\documentclass[11pt]{amsart}
\usepackage{amsfonts}
\usepackage{amsmath}
\usepackage{amssymb}
\usepackage{cite}
\usepackage{color}
\usepackage{graphicx}
\usepackage{xspace}
\usepackage{axodraw}
\usepackage{bbm}

\newtheorem{prop}{Proposition}[section]
\newtheorem{lem}{Lemma} [section]
\newtheorem{cor}{Corollary}[section]

\newtheorem{thm}{Theorem}[section]

\newcommand{\fish}{\parbox{3pc}{\begin{picture}(20,20)
\put(4,10){\qbezier(0,-4)(16,20)(32,-4)}
\put(4,10){\qbezier(0,4)(16,-20)(32,4)}
\end{picture}}}

\DeclareMathOperator{\Char}{Char}  
\DeclareMathOperator{\End}{End}    
\DeclareMathOperator{\Lin}{Lin}    
\DeclareMathOperator{\Res}{Res}    
\newcommand{\vf}{\varphi}          
\newcommand{\eps}{\varepsilon}     
\newcommand{\CC}{\mathbb{C}}       
\newcommand{\N}{\mathbb{N}}        
\newcommand{\Q}{\mathbb{Q}}        
\newcommand{\R}{\mathbb{R}}        
\newcommand{\cS}{\mathcal{S}}      
\newcommand{\Z}{\mathbb{Z}}        

\newcommand{\barast}{\mathbin{\overline\ast}} 
\DeclareMathOperator{\Lie}{Lie}    
\DeclareMathOperator{\BCH}{BCH}    

\newcommand{\id}{\mathrm{id}}

\newcommand{\sepword}[1]{\quad\mbox{#1}\quad}

\newcommand{\hideqed}{\renewcommand{\qed}{}}

\begin{document}

\title{A Lie theoretic approach to renormalization}

\author{Kurusch Ebrahimi-Fard}
\address{I.H.\'E.S.,
         Le Bois-Marie,
         35, Route de Chartres,
         F-91440 Bures-sur-Yvette, France}
\email{kurusch@ihes.fr}
\urladdr{http://www.th.physik.uni-bonn.de/th/People/fard/}

\author{Jos\'e M. Gracia-Bond\'{\i}a}
\address{Departamento de F\'{\i}sica Te\'orica I,
         Universidad Complutense,
         Madrid 28040, Spain\
         {\rm and} \newline\null\indent
         Departamento de F\'{\i}sica,
         Universidad de Costa Rica,
         San Pedro 2060, Costa Rica}

\author{Fr\'ed\'eric Patras}
\address{Laboratoire J.-A. Dieudonn\'e
         UMR 6621, CNRS,
         Parc Valrose,
         06108 Nice Cedex 02, France}
\email{patras@math.unice.fr}
\urladdr{www-math.unice.fr/~patras}


\begin{abstract}

Motivated by recent work of Connes and Marcolli, based on the
Connes--Kreimer approach to renormalization, we augment the latter by
a combinatorial, Lie algebraic point of view.  Our results rely both
on the properties of the Dynkin idempotent, one of the fundamental Lie
idempotents in the theory of free Lie algebras, and on properties of
Hopf algebras encapsulated in the notion of associated descent
algebras.  Besides leading very directly to proofs of the main
combinatorial aspects of the renormalization procedures, the new
techniques give rise to an algebraic approach to the Galois theory of
renormalization.  In particular, they do not depend on the geometry
underlying the case of dimensional regularization and the
Riemann--Hilbert correspondence.  This is illustrated with a
discussion of the BPHZ renormalization scheme.

\end{abstract}

\maketitle


\tableofcontents

\thispagestyle{empty}


\section{Introduction}
\label{sect:introduction}

{}From its inception, renormalization theory in perturbative quantum
field theory (pQFT) had a combinatorial flavour, as well as an
analytic one. The former manifests itself in the self-similarity of
Feynman graphs, the building blocks of pQFT. The intricate
combinatorics of extracting and combining subgraphs, required in the
renormalization process, is encoded in the Bogoliubov recursion,
respectively its solution via Zimmermann's forest
formula~\cite{CaswellK1982,collins1984,Lowen1975,Vasilev04,Zimmermann}.

Kreimer's discovery of a Hopf algebra structure underlying Bogoliubov
and Zimmermann's formulae and illuminating their internal
structure~\cite{kreimer1998} was the starting point of a new approach
in the field.  Then the Connes--Kreimer decomposition \`{a} la
Birkhoff--Wiener--Hopf (BWH) of Feynman rules~\cite{ck2000} captured
the process of renormalization in pQFT in the framework of dimensional
regularization (DR) with the minimal subtraction (MS) scheme.  Further
work by Connes, Kreimer and others has since then established various
links between the BWH decomposition of characters in renormalizable
quantum field theories and relevant mathematical topics, culminating
recently in work by Connes and Marcolli on motivic Galois
theory~\cite{cm22004,cm2006} and by Bloch, Esnault and Kreimer on the
motivic nature of primitive Feynman graphs~\cite{BEK2005}.

In the present work, largely motivated by~\cite{cm22004}, we return to
the origin of the Connes--Kreimer theory and concentrate on algebraic
features of renormalization relevant to pQFT, trying to unravel
further fundamental properties of renormalization schemes by methods
inspired on the classical theory of free Lie algebras (FLAs).

It has been known since the mid-nineties that many properties of~FLAs,
as exposed e.g. in~\cite{OldNiko89,reutenauer1993}, can be lifted to
general graded Lie algebras and their enveloping algebras.  In other
terms Lie theory is relevant to the study of arbitrary graded
connected cocommutative or commutative Hopf algebras.  In particular,
the Solomon algebras of type $A_n$~\cite[Chap.~9]{reutenauer1993} act
naturally on arbitrary graded connected commutative Hopf
algebras~\cite[Thm.~II.7]{patras1994}.

The observation applies to the Hopf algebras of renormalization,
yet it has not received the attention it deserves.  Here we
develop it systematically, considering abstract renormalization
Hopf algebras~$H$ and commutative target algebras~$A$ of quantum
amplitudes endowed with a Rota--Baxter operator.  We show that
some of the deepest combinatorial properties of renormalization
schemes are codified in the composition with the Dynkin idempotent
of Lie theory, and in its inverse map.  We derive in particular
from their study the properties of characters under locality
assumptions for counterterms in renormalization and, in the
process, establish that the data relevant to their computation are
contained in the ``beta function''.  The phenomenon is well known
in pQFT; the Lie theoretic approach, however, provides a
remarkably efficient way to deduce it from the locality
assumptions.

Furthermore, the direct sum of Solomon algebras (the
\textit{descent algebra}) is naturally provided with a graded
connected cocommutative Hopf algebra structure; the corresponding
pro-unipotent group is naturally isomorphic to the universal
group~$U$ of the Connes--Marcolli Galois theory of
renormalization.  This isomorphism paves the way to an algebraic
and combinatorial approach to the later theory.
Some advantages of our method are: (i)~It appears to be independent of
the DR with MS prescription, applying in principle to any
renormalization procedure which can be formulated in terms of a
Rota--Baxter structure; (ii)~Use of the Dynkin map explains naturally
the coefficients of the universal singular frame of~\cite{cm22004}
---the same as in the Connes--Moscovici index formula~\cite{gafa95}
for the case of simple dimension spectrum.

The article is organized as follows.  After settling some notation in
the next section, we ponder in Section~\ref{sect:HopfCharacters} the
convolution algebra of linear maps~$\Lin(H,A)$.  It cannot be made
into a Hopf algebra in general; but a suitable algebra of
\textit{characteristic functions} can.  This is our playground; it
encodes, at the Hopf algebra level, the properties of the group of
$A$-valued characters of $H$.  Section~\ref{sect:dynkinOperator} is
the heart of the paper: starting from a short survey on the Dynkin
idempotent for cocommutative Hopf algebras, we establish the formal
properties of its sibling in the commutative case, then introduce and
exhaustively study the inverse Dynkin map.  In particular, we show
that the latter bijectively sends infinitesimal characters into
characters of commutative connected Hopf algebras ---applying in
particular to the Hopf algebras of Feynman diagrams and rooted trees
of renormalization theory and the corresponding Feynman rules.  In
Section~\ref{sect:birkhoffdecomp} we recall the BWH decomposition of
characters, choosing once again to obtain it algebraically from
Rota--Baxter operator theory and the `Baker--Campbell--Hausdorff (BCH)
recursion'.  After that, our Lie theoretic machine is fully
operational.

In the rest of the paper, we show the relevance of that machine to
pQFT. In Section~\ref{sect:reminder} we briefly review some
standard lore of renormalization and remind the reader of the
dictionary between it and the Connes--Kreimer paradigm.  Next we
study in Section~\ref{sect:locality} the locality properties for
dimensional regularization (DR) schemes by exploiting the
properties of the Dynkin pair of maps, together with the BWH
decomposition.  The main results concerning the Connes--Kreimer
beta function and the renormalization group (RG) in DR are
rederived from direct computations in the algebra of
characteristic functions; hopefully, the role of that beta
function is thereby illuminated.  Sections~\ref{sect:DSlocality}
and~\ref{sect:BPHZlocality} are essays on the same question in
other renormalization frameworks; in the second we invoke the BPHZ
scheme of renormalization and exhibit the underlying Rota--Baxter
algebra structure, exemplifying with the
(Ginzburg--Landau--Wilson) $\vf^4_4$~scalar model in Euclidean
field theory.

To finish, in Section~\ref{sect:cosmic} we go back to the mathematical
setting, trying to place our results in the `great scheme of things'
of combinatorial Hopf algebra theory.  We show there how the
Connes--Marcolli ``motivic Galois group'' of renormalization relates
with FLAs as well as the theory of descent algebras.  Together with
the links between the same group and Connes--Moscovici's index theorem
in noncommutative geometry, these new connections give further
evidence for Connes' and Kreimer's ---already much documented--- claim
that the divergences of pQFT do reveal the presence of deep
mathematical structures.


\section{Notational conventions}
\label{sect:convs}

Let $H=\bigoplus_{n=0}^{\infty}H_n$ be a graded connected commutative
Hopf algebra (of finite type) over a field~$k$ of characteristic zero;
this is necessarily free as a commutative
algebra~\cite[Prop.~4.3]{patras1994}.  We write~$\epsilon$ for the
augmentation from~$H$ to~$H_0=k\subset H$ and~$H^+$ for the
augmentation ideal $\bigoplus_{n=1}^{\infty} H_n$ of~$H$.  The
identity map of~$H$ is denoted~$I$.  The product in~$H$ is
written~$\pi$ or simply by concatenation.  The coproduct is
written~$\delta$; we use Sweedler's notation and write $h^{(1)}\otimes
h^{(2)}$ for~$\delta (h),\,h\in H_n$; or $\sum_{i=0}^n
h_i^{(1)}\otimes h_{n-i}^{(2)}$ when the grading has to be taken into
account.  The usual restrictions apply, that is, $h^{(1)}\otimes
h^{(2)}$ stands for a sum $\sum_{j\in J}h_j^{(1)}\otimes h_j^{(2)}$
and should not be interpreted as the tensor product of two elements
of~$H$.  The same convention applies in forthcoming notation such as
$h^{(1)}\otimes g^{(1)}\otimes h^{(2)}\otimes g^{(2)}$, that should be
understood as $\sum_{j\in J}\sum_{k\in K}h_j^{(1)}\otimes
g_k^{(1)}\otimes h_j^{(2)}\otimes g_k^{(2)}$, where $\delta
(g)=\sum_{k\in K}g_k^{(1)}\otimes g_k^{(2)}$.

Graduation phenomena are essential for all our forthcoming
computations, since in the examples of physical interest they
incorporate information such as the number of loops (or vertices) in
Feynman graphs, and the subdivergence structure, relevant for the~RG.
They are expressed by the action of the grading operator $Y:H\to H$,
given by:
$$
    Y(h) = \sum\limits_{n\in\N} n h_n \sepword{for} h =
    \sum\limits_{n\in\N}h_n\in\bigoplus\limits_{n\in\N} H_n.
$$
We write $|h_n|:=n$. Notice that $Y$ is usually denoted by~$D$ in the
FLA literature. In the present article we stick to the notation most
often used in the context of the Hopf algebra approach to
renormalization~\cite{ck1998,ck2000,ck2001,cm22004,ek2005,fg2005} and
reserve~$D$ for the Dynkin operator.


\section{The Hopf algebra of characteristic functions}
\label{sect:HopfCharacters}

A \textit{character} is a map $\gamma$ of unital algebras
from~$H$ to the base field $k$:
$$
    \gamma (hh') = \gamma(h)\gamma(h').
$$
It should be clear that the product on the right hand side is the one
in~$k$. We write $\gamma_n$ for the restriction of $\gamma$ to a map
from~$H_n$ to~$k$.

Let $A$ be a commutative $k$-algebra, with unit~$1_A=\eta_A(1)$,
$\eta_A:k\to A$ and with product~$\pi_A$, which we sometimes denote by
a dot: $\pi_A(u\otimes v)=:u\cdot v$. The main examples we have in
mind are $A=\CC,\,A=\CC[[\varepsilon,\varepsilon^{-1}]$ and~$A=H$. We
extend now the definition of characters and call an ($A$-valued)
character of $H$ any algebra map from~$H$ to~$A$. In particular
$H$-valued characters are simply algebra endomorphisms of $H$.

An \textit{infinitesimal character} is a linear map~$\alpha$
from~$H$ to~$k$ such that:
$$
    \alpha (h h') = \alpha(h) \epsilon(h') + \epsilon (h) \alpha (h').
$$
As for characters, we write $\alpha(h) = \sum_{n\in\N} \alpha_n
(h_n)$. The same notational convention will be used in the sequel
without further notice: $f_n$ stands for the restriction of an
arbitrary map~$f$ on~$H$ to $H_n$. We remark that by the very
definition of characters and infinitesimal characters
$\gamma_0(1_H)=1$, that is $\gamma_0 = \epsilon$, whereas
$\alpha_0(1_H)=0$.

We extend as well in the obvious way the notion of infinitesimal
characters to maps from~$H$ to a commutative $k$-algebra $A$. We
have now:
$$
    \alpha(hh') = \alpha(h) \cdot e(h') + e(h) \cdot \alpha (h'),
$$
where $e:=\eta_A\circ\epsilon$. They can be alternatively defined
as $k$-linear maps from $H$ to $A$ with $\alpha_0=0$ that vanish
on the square of the augmentation ideal of $H$. In particular, if
$\alpha$ is an infinitesimal character, the linear map
$\alpha^{|n}$, the restriction of which is~0 on all the graded
components of~$H$ excepted $H_n$, and $\alpha^{|n}_n= \alpha_n$,
is also an infinitesimal character. Thus $\alpha$ decomposes as a
sum of infinitesimal characters $\alpha =\sum_{n>0}\alpha^{|n}$.
The vector space of infinitesimal characters, written $\Xi_H(A)$,
or just $\Xi(A)$, decomposes accordingly as the direct product of
its graded components: $\Xi(A)=\prod_{n\in\N^\ast} \Xi_n(A)$,
where~$\Xi_n(A)$ is the linear span of the~$\alpha^{|n}$. Thus we
regard $\Xi(A)$ as the natural `topological' completion of the
graded vector space $\oplus_{n\in\N^\ast}\Xi_n(A)$. In more
detail, we consider the subspaces $\oplus_{i\ge n\in\N^\ast}
\Xi_i(A)$ and the associated onto homomorphisms, and
we take the inverse limit, whose elements are infinite series.
This property we agree to abbreviate from now on to ``$\Xi(A)$ is
a graded vector space''; the sundry objects defined below are
graded in that sense ---that is, they are actually completions of
graded vector spaces, completions of graded Lie algebras, and so
on.

The space $\Lin(H,A)$ of $k$-linear maps from~$H$ to~$A$,
$\Lin(H,A):=\prod_{n\in\N}\Lin(H_n,A)$, is naturally endowed with
an algebra structure by the \textit{convolution product}:
 \allowdisplaybreaks{
\begin{equation*}
    f\ast g := \pi_A\circ(f\otimes g)\circ\delta: \qquad H
    \xrightarrow{\delta} H \otimes H \xrightarrow{f \otimes g} A
    \otimes A \xrightarrow{\pi_{A}} A.
\end{equation*}}
The unit for the convolution product is precisely~$e: H \to A$.
Especially when $A=H$, the convolution product provides various tools
to deal with properties of characters, such as the logarithm of the
identity, which is a projector on~$H$ with kernel the square of the
augmentation ideal. As a corollary, $A$-valued infinitesimal
characters can be characterized as those maps~$\alpha$ from~$H$ to~$A$
such that $\alpha_n \circ I^{\ast k}(h_n)=k \,\alpha_n(h_n)$, for
any~$k,n\in\N$. We refer the reader interested in a systematic study
of these phenomena and of their connections to the classical structure
theorems for Hopf algebras such as the Cartier-Milnor-Moore theorem
to~\cite{patras1993,patras1994,Cartier2006}.

The set $G_H(A)$ of characters, or just $G(A)$, is a group for the
convolution product. The inverse is given explicitly by the
formula:
\begin{equation*}
                \gamma^{-1} = \Big(e + \sum\limits_{n\in\N^\ast}
                \gamma_n\Big)^{-1} = e +
                \sum\limits_{k\in\N^\ast}(-1)^{k}\,
                \bigg(\,\sum\limits_{n\in\N^\ast}
                \gamma_n\bigg)^{\ast k}.
\end{equation*}
The last sum is well-defined as a power series, since only a
finite number of terms appear in each degree. We denote as usual
by~$S$ the convolution inverse of the identity map~$I$
in~$\End(H):=\Lin(H,H)$. Then $\gamma^{-1}= \gamma \circ S \in
G(A)$; the reader unfamiliar with this identity can deduce it
easily from the next lemma and other notions introduced soon.

Now, $\Lin(H,A)$ is \textit{not} naturally provided with a Hopf
algebra structure over the ground field $k$, except under
particular assumptions on the target space $A$. For example, it is
(up to the completion phenomena) a Hopf algebra if $A=k$. This
follows from the usual argument to construct a Hopf algebra
structure on the graded dual of a graded connected Hopf algebra of
finite type. It is not when $A=k[[\varepsilon,\varepsilon^{-1}]$,
that is when the coefficient algebra~$A$ is the field of Laurent
series ---an example relevant to renormalization. However, as will
be shown below, a Hopf algebra structure can always be defined on
certain remarkable spaces naturally related to~$\Lin(H,A)$ and,
most importantly in view of applications to pQFT, to the group of
characters $G(A)$.

\begin{lem}
\label{lem:AstCirc}
Assume that, for given $\phi ,\psi \in \Lin(H,A)$, there exist
elements $\phi^{(1)}\otimes\phi^{(2)}$, respectively
$\psi^{(1)}\otimes \psi^{(2)}$, in $\Lin(H,A)\otimes \Lin(H,A)$ such
that, for any $h,h'\in H$\/:
$$
    \phi^{(1)}(h) \cdot \phi^{(2)}(h') = \phi (hh') \sepword{and}
    \psi^{(1)}(h) \cdot \psi^{(2)}(h') = \psi (hh');
$$
then
$$
                            \phi \ast \psi (h h') = \big(\phi^{(1)}
                            \ast \psi^{(1)}(h)\big) \cdot
                            \big(\phi^{(2)} \ast \psi^{(2)}(h')\big).
$$
Moreover, when $\psi\in \End(H)$ and $\phi\in \Lin(H,A)$, with the
same hypothesis and $\psi^{(1)},\psi^{(2)}$ now in $\End(H)$\/:
$$
                                 \phi \circ \psi (hh') =
                                 \big(\phi^{(1)} \circ
                                 \psi^{(1)}(h)\big) \cdot
                                 \big(\phi^{(2)} \circ
                                 \psi^{(2)}(h')\big).
$$
The last identity in particular holds when $A=H$, that is,
in~$\End(H)$.
\end{lem}
\begin{proof}
Indeed, we have:
 \allowdisplaybreaks{
\begin{eqnarray*}
    \phi \ast \psi (hh') &=& \phi(h^{(1)}{h'}^{(1)}) \cdot
                             \psi(h^{(2)}{h'}^{(2)})\\
                         &=& \phi^{(1)}(h^{(1)}) \cdot
                             \phi^{(2)}({h'}^{(1)}) \cdot
                             \psi^{(1)}(h^{(2)}) \cdot
                             \psi^{(2)}({h'}^{(2)})\\
                         &=& \phi^{(1)}(h^{(1)}) \cdot
                             \psi^{(1)}(h^{(2)}) \cdot
                             \phi^{(2)}({h'}^{(1)}) \cdot
                             \psi^{(2)}({h'}^{(2)})\\
                             &=& \big(\phi^{(1)}\ast\psi^{(1)}(h)\big)
                             \cdot
                             \big(\phi^{(2)}\ast\psi^{(2)}(h')\big),
\end{eqnarray*}}
an identity that we also write, for later use,
$$
    (\phi\ast\psi )^{(1)}\otimes (\phi\ast\psi )^{(2)}=
    (\phi^{(1)}\otimes \phi^{(2)})\ast (\psi^{(1)}\otimes \psi^{(2)}).
$$
We also clearly have:
$$
    \phi \circ \psi (hh') = \phi (\psi^{(1)}(h)\ \psi^{(2)}(h'))
                          = \big(\phi^{(1)}\circ\psi^{(1)}(h)\big) \cdot
                            \big(\phi^{(2)} \circ \psi^{(2)}(h')\big).
    \eqno \qed
$$
\hideqed
\end{proof}

\begin{cor}
The graded vector space of infinitesimal characters $\Xi(A)$ is a
graded Lie subalgebra of~$\Lin(H,A)$ for the Lie bracket induced
on the latter by the convolution product.
\end{cor}
\begin{proof}
Indeed, infinitesimal characters are precisely the elements
$\alpha$ of~$\Lin(H,A)$ such that:
$$
\alpha^{(1)}\otimes\alpha^{(2)} = \alpha\otimes e + e\otimes\alpha
\sepword{satisfy, for any $h,h'\in H$:} \alpha^{(1)}(h) \cdot
\alpha^{(2)}(h') = \alpha(hh').
$$
According to the foregoing lemma, for $\alpha$ and $\beta$ two graded
infinitesimal characters, we obtain:
 \allowdisplaybreaks{
\begin{eqnarray}
    [\alpha ,\beta ](hh') &:=& (\alpha\ast\beta -\beta\ast\alpha)(hh')
    \nonumber\\
    &=& \pi_A [ (\alpha\otimes e + e \otimes \alpha)\ast(\beta
    \otimes e + e \otimes \beta)
    \nonumber\\
    & & -(\beta \otimes e + e \otimes \beta) \ast (\alpha\otimes e + e
    \otimes \alpha)](h \otimes h')
    \nonumber\\
    &=& [\alpha ,\beta ](h)\cdot e(h') + e (h)\cdot [\alpha,\beta](h'),
    \nonumber
\end{eqnarray}}
hence the corollary follows.
\end{proof}

\begin{prop}
The enveloping algebra $U(\Xi (A))$ of the Lie algebra~$\Xi(A)$ maps
naturally to the convolution subalgebra of~$\Lin(H,A)$ generated
by~$\Xi (A)$.
\end{prop}

The existence of that natural algebra map from~$U(\Xi(A))$ to~$\Lin(H,A)$
follows from the previous lemma and from the universal property of
enveloping algebras.

Notice that $U(\Xi(A))$ is also, as the enveloping algebra of a graded
Lie algebra, a graded connected cocommutative Hopf algebra, which we
call the Hopf algebra~$\Char_H(A)$, or just~$\Char(A)$, of
characteristic functions on~$H$ (with values in~$A$). We write
$\barast$ for the product on~$\Char(A)$ and use~$\Delta$ for its
coproduct. Thus by definition of~$\Char(A)$ the primitive elements are
the infinitesimal $A$-valued characters of~$H$. Besides providing Hopf
algebra tools for the study of Feynman rules, the Hopf algebra of
characteristic functions ---and the associated pro-unipotent group---
will play a crucial role in Section~\ref{sect:cosmic}, when relating
the FLA approach to renormalization to the Connes--Marcolli motivic
Galois group.

Notice that $\Delta$ is not defined in general on $\Lin(H,A)$, and
neither on the image of $\Char(A)$ in $\Lin(H,A)$, see~\cite{hazy04}
and~\cite{patreu2002} for a discussion on the subject in the
particular case $A=H$.

\begin{prop}
\label{prop:recilaw}
We have, for any $\phi \in \Char(A)$ and any $h,h'\in H$, the
reciprocity law:
$$
    \phi (hh') = \phi^{(1)}(h) \cdot \phi^{(2)}(h'),
$$
where we use the Sweedler notation for~$\Delta(\phi)$, and where the
action of~$\phi$ on~$H$ is induced by the natural map from~$\Char(A)$
to~$\Lin(H,A)$.
\end{prop}
\begin{proof}
This is true when~$\phi$ is an infinitesimal character. According
to the previous proposition, for $\phi,\phi '$ in $\Char(A)$, we
have: $\phi\barast\phi'(hh')=\phi\ast\phi' (hh')$. Due to
the Lemma \ref{lem:AstCirc}, it follows that the identity holds
for $\phi\barast\phi'$ if it holds for $\phi$ and $\phi'$.
Since $\Char(A)$ is generated as an associative algebra by
infinitesimal characters, the proposition follows.
\end{proof}

We remark that the reciprocity law can be rewritten:
$$
    \phi \circ \pi = \pi_A \circ \Delta(\phi).
$$
In the cocommutative case, the identity playing a similar role
(mutatis mutandis) is~\cite{patreu2002}:
$$
    \delta \circ \phi  = \Delta(\phi) \circ \delta.
$$

\smallskip

As a consequence of Proposition~\ref{prop:recilaw}, the set
$G'(A)$ of group-like elements in $\Char(A)$ maps to characters,
that is elements of $G(A)$ ---since the identity $\Delta
(\phi)=\phi\otimes\phi$ in $\Char(A)$ translates into the identity
$\phi(hh')=\phi(h)\phi(h')$ in $H$. We show now that, as usual,
the convolution exponential and logarithm maps are inverse
bijections from $\Xi (A)$ onto~$G(A)$ and from $\Xi(A)$ onto
$G'(A)$. Indeed, for any $\alpha\in\Xi (A)$, we have in
$\Char(A)$:
 \allowdisplaybreaks{
\begin{eqnarray*}
    \Delta \big(\exp(\alpha)\big) &=& \exp\big(\Delta (\alpha) \big)
    =  \exp(\alpha\otimes e + e \otimes\alpha )
    =  \exp (\alpha\otimes e) \barast \exp(e \otimes\alpha) \\
    &=& \big(\exp (\alpha)\otimes e \big) \barast \big(e\otimes
    \exp(\alpha)\big) = \exp(\alpha) \otimes \exp(\alpha),
\end{eqnarray*}}
which also implies that we have $\exp(\alpha )\in G(A)$
in~$\Lin(H,A)$. We have used first that $\alpha$ is a graded
infinitesimal character, then that $\alpha\otimes e$ and
$e\otimes\alpha$ commute. The other way round, if $\gamma$ is a
character:
 \allowdisplaybreaks{
\begin{eqnarray*}
     \log(\gamma )(hh') &=& \pi_A\big( \log(\gamma\otimes\gamma)(h
                        \otimes h') \big) \\
                        &=& \pi_A\big((\log(\gamma\otimes e) +
                        \log(e\otimes\gamma))(h\otimes h')\big) \\
                        &=& \pi_A\big((\log(\gamma)\otimes e + e
                        \otimes \log(\gamma))(h\otimes h')\big) \\
                        &=& \log(\gamma)(h) \cdot e(h') + e(h) \cdot
                        \log(\gamma)(h'),
\end{eqnarray*}}
whereas if $\gamma\in G'(A)$:
 \allowdisplaybreaks{
\begin{eqnarray*}
     \Delta (\log \gamma) &=& \log (\Delta \gamma ) =
     \log (\gamma\otimes\gamma ) \\
      &=& \log ((\gamma\otimes e)\barast (e\otimes\gamma )) =
      \log\gamma\otimes e + e\otimes\log\gamma .
\end{eqnarray*}}

\begin{cor}
The natural algebra map from~$\Char(A)$ to~$\Lin(H,A)$ induces a
bijection between the group $G'(A)$ of group-like elements
in~$\Char(A)$ and $G(A)$, the group of $A$-valued characters on~$H$.
\end{cor}

We identify $G(A)$ with~$G'(A)$ henceforth. In particular, both the
identity map~$I$ and the antipode~$S$ can be viewed as elements
of~$\Char(H)$, and we won't distinguish between~$I,S$ and their
respective preimages in~$\Char(H)$.


\section{Logarithmic derivatives and the Dynkin operator}
\label{sect:dynkinOperator}

Although the logarithm is the simplest bijection from group-like
elements to primitive elements in a graded connected cocommutative
Hopf algebra, the most relevant bijection in view of applications to
renormalization is a bit subtler. It is a kind of logarithmic
derivative, closely related to a Lie idempotent known as the
\textit{Dynkin idempotent}. Presently we survey the properties of the
Dynkin operator (the composition of the Dynkin idempotent with the
grading map $Y$) pertinent to our forthcoming computations, and also
obtain new results on the operator, such as the existence of the
advertised bijection between~$G(A)$ and~$\Xi(A)$. The results
generalize the fine properties of Hopf algebras encapsulated in the
notion of descent algebra of a Hopf algebra~\cite{patras1994}. They
rely as well on the Hopf algebraic treatment of the Dynkin operator
given in~\cite{patreu2002}, and on more classical Lie theoretic
properties of the operator. We give in particular closed formulas for
the inverse map to Dynkin's operator, i.e., from $\Xi(A)$ to $G(A)$.

\smallskip

The classical Dynkin operator is defined as follows. Let $X=
\{x_1,\ldots,x_n,\ldots\}$ be an alphabet. The tensor algebra
$T(X):=\bigoplus_{n\geq 0}T_n(X)$ over~$X$ is a cocommutative graded
Hopf algebra with the set of words $x_{i_1}\ldots x_{i_l}$ as a linear
basis in degree $l$. It is also, canonically, the enveloping algebra
of the FLA~$\Lie(X)$ over~$X$. The product in~$T(X)$ is induced by
concatenation of words, whereas the coproduct is fully characterized
by the property that elements of~$X$ are primitive in~$T(X)$. The
Dynkin operator $D:T(X)\to\Lie(X)$ is given by:
$$
    D(x_{i_1}\ldots x_{i_n}) =
    [\dots[[x_{i_1},x_{i_2}],x_{i_3}],\dots,x_{i_n}];
$$
with $D_{T_0(X)}=0$ and $D_{T_1(X)}=\id_X$. According to an idea
essentially due to von~{\makebox{Waldenfels}}, this iterated
bracketing operator can be represented in a more abstract way as
the convolution product of the antipode~$S$ of~$T(X)$ with the
grading operator~$Y$, acting as the multiplication by~$n$
on~$T_n(X)$:
\begin{equation*}
    D = S \ast Y; \sepword{equivalently} I \ast D = Y.
\end{equation*}
The most famous property of~$D$ is the Dynkin--Specht--Wever theorem,
stating that an element~$v$ in $T_n(X)$ ---a linear combination of
words of length $n$--- is in~$\Lie(X)$ if and only if $D(v) = nv$. In
effect, such a~$v$ is in the primitive part of the tensor algebra,
and:
$$
    nv = Yv = \pi(I \otimes D)(1 \otimes v + v \otimes 1) = D(v).
$$
The converse is also well known.  These definitions and properties
have been generalized to bialgebras in~\cite{patreu2002}, that we
follow mainly here.  However, for our purposes we need to give a
somewhat detailed account.  Indeed, that reference as well as the
classical theory of the Dynkin operator do focus on the study of
graded connected cocommutative Hopf algebras, whereas we are mainly
interested here in \textit{commutative} Hopf algebras.  The interested
reader will find further historical and technical information about
the Dynkin operator and its relevance to Lie computations, as well as
other classical Lie idempotents, in references~\cite{gelfand1995}
and~\cite{reutenauer1993}.

So let $H$ be graded, connected and commutative. Since $I\in\Char(H)$,
so its graded components $I_n\in\Char(H)$. Notice, for further use,
that the subalgebra of $\Char(H)$ generated by the~$I_n$ maps to the
descent algebra of~$H$ ---the convolution subalgebra of~$\End(H)$
generated by the~$I_n$, see \cite{patras1994}. Moreover, the
grading operator $Y:= \sum_{n\in\N} n I_n$ belongs to~$\Char (H)$.
Its coproduct is given by:
$$
    \Delta (Y) = Y \otimes I + I \otimes Y,
$$
another way of expressing that $Y$ is a derivation of~$H$:
$$
    Y(hh') = Y(h)h'+ hY(h').
$$
Let us adopt the notation $Yf$ for $f\circ Y$, according to the custom
in distribution theory. Under this guise the operator $Y$ extends
naturally to a derivation on $\Lin(H,A)$. We find,
with~$f,g\in\Lin(H,A)$ and $h\in H$:
 \allowdisplaybreaks{
\begin{eqnarray*}
    Y(f \ast g)(h) &:=& f \ast g\,(Y(h)) = |h|(f \ast g)(h) \\
                   &=& |h|f(h^{(1)})g(h^{(2)}) \\
                   &=& |h^{(1)}|f(h^{(1)})g(h^{(2)}) +
                   |h^{(2)}|f(h^{(1)})g(h^{(2)}) \\
                   &=& Yf \ast g\,(h) + f \ast Yg\,(h),
\end{eqnarray*}}
where we used that $\Delta (Y(h))=|h|\Delta (h)=
\big(|h^{(1)}|+|h^{(2)}|\big)\,h^{(1)}\otimes h^{(2)}$.

\begin{prop} \label{prop:Sconvd}
Convolution of the $H$-valued character~$S$ with any derivation~$d$
of~$H$ yields an $H$-valued infinitesimal character.
\end{prop}

\begin{proof}
Indeed, since $d$ is a derivation, we have $d(hh')=d(h)h'+hd(h')$.
Since, furthermore, $\Delta (S)=S\otimes S$, we get:
 \allowdisplaybreaks{
\begin{eqnarray*}
(S \ast d)(hh') &=& \pi\circ [ (S\otimes S) \ast (d\otimes I+I\otimes d)]
                (h\otimes h') \\
                &=&\pi\circ [ (S \ast d) \otimes (S \ast I)
                + (S \ast I) \otimes (S \ast d)](h\otimes h') \\
                &=& S \ast d (h)\cdot e(h') + e(h) \cdot S
                \ast d(h'),
\end{eqnarray*}}
hence the proposition follows.
\end{proof}

\begin{cor} \label{cor:DynkinInfChar}
The Dynkin operator $D:=S\ast Y$ is an $H$-valued infinitesimal
character of~$H$.
\end{cor}

Notice also that $D$ satisfies $D\circ D=D\circ Y$ ---in other terms,
$D$ is an idempotent up to a scalar factor depending on the grading,
that is, $D$ is a quasi-idempotent. Equivalently, $D\circ Y^{-1}$ is
an idempotent on $H^+$ (also known when $H$ is the cotensor algebra
$T^\ast(X)$ ---see below--- as the Dynkin idempotent). Indeed, for any
$h\in H$,
$$
    D\circ D(h) = D\circ (S\ast Y)(h)
                = D\big(S(h^{(1)}) Y(h^{(2)})\big).
$$
However, since $D$ is an infinitesimal character, $D(hh')=0$ if
$h,h' \in H^+$ and therefore,
$$
    D\circ D(h) = D\big(S(h)Y(1_H)+S(1_H)Y(h)\big)
                = D\circ Y(h),
$$
since $Y(1_H)=0$.

\begin{prop}
Right composition with an infinitesimal
character~$\alpha\in\Xi(H)$ induces a map from~$G(H)$ to~$\Xi(H)$.
This also holds for~$G(A)$ and~$\Xi(A)$, where~$A$ is an arbitrary
commutative unital algebra.
\end{prop}

\begin{proof}
Indeed, let $\gamma \in G(H)$ or $G(A)$, we have:
$$
    \gamma \circ \alpha (hh') = \gamma\circ\alpha(h)\,e(h') +
    e(h)\,\gamma\circ\alpha(h'),
$$
by virtue of Lemma~\ref{lem:AstCirc}, since $\gamma\circ e=e$ for
any character~$\gamma$.
\end{proof}

\begin{cor}
\label{cor:DynkinD}
Right composition with the Dynkin operator $D$ induces a map
from~$G(A)$ to~$\Xi(A)$.
\end{cor}

In general, for~$\gamma$ belonging to~$G(H)$ or~$G(A)$ and any
$f_1,\ldots,f_k\in\End(H)$, we have the distributive property:
$$
    \gamma \circ (f_1 \ast \dots \ast f_k) = (\gamma\circ f_1) \ast
    \dots \ast (\gamma\circ f_k).
$$
Particularly,
$$
    \gamma \circ D = \gamma \circ (S \ast Y)
                   = \gamma^{-1} \ast Y\gamma.
$$

\begin{thm}
\label{thm:Gamma}
Right composition with $D$ is a bijective map from~$G(H)$ to~$\Xi(H)$.
The inverse map is given by:
\begin{equation}
    \Gamma : \alpha\in\Xi (H) \longmapsto \sum\limits_n
    \sum_{\substack{k_1, \dots, k_l\in\N^\ast \\ k_1 + \dots +
      k_l = n}} \frac{\alpha_{k_1} \ast
    \dots\ast \alpha_{k_l}} {k_1(k_1+k_2) \dots (k_1+ \dots +k_l)} \in
    G(H).
\label{eq:madre-del-cordero}
\end{equation}
The theorem also holds if $G(H)$ and~$\Xi(H)$ are replaced
by~$G(A)$, respectively~$\Xi(A)$.
\end{thm}

We show first that~$\Gamma$ is a left inverse to right composition
with~$D$. The following lemma has been obtained in~\cite{gelfand1995}
in the setting of noncommutative symmetric functions and
quasi-determinants. We include the proof, which is elementary.

\begin{lem}\label{dynkid}
For~$n\ge1$ we have:
$$
I_n = \sum_{\substack{k_1, \dots, k_l\in\N^\ast \\ k_1 + \dots +
      k_l = n}} \frac{D_{k_1}\ast \dots \ast D_{k_l}}{k_1(k_1 + k_2)
      \dots (k_1 + \cdots + k_l)}.
$$
\label{lem:lemma-of-lemmata}
\end{lem}

\begin{proof}
As already remarked, the definition of~$D$ implies $I \ast D = I
\ast S \ast Y = Y$. In particular, since $D_0=0$:
$$
    Y_n = n I_n = (I \ast D)_n = \sum\limits_{i=1}^{n}I_{n-i}\ast D_i.
$$
Inserting recursively the value of~$I_i$ in the right member of
the identity, we obtain:
 \allowdisplaybreaks{
\begin{eqnarray*}
    I_n &=& \frac{D_n}{n} + \sum\limits_{i=1}^{n-1}\sum\limits_{1\leq
    j\leq n-i} \frac{I_{n-i-j}\ast D_{j} \ast D_i}{(n-i)n}
    \\
    &=& \frac{D_n}{n} + \sum_{\substack{j+i=n \\ i,j\not=0}}
      \frac{D_j\ast D_i}{j\cdot n} + \sum\limits_{i=1}^{n-1}
    \sum\limits_{1\leq j\leq n-i-1}\frac{I_{n-i-j}\ast D_{j} \ast
    D_i}{(n-i)n}
    \\
    &=& \sum_{\substack{k_1,\dots,k_l\in \N^\ast \\ k_1 + \dots +
    k_l=n}} \frac{D_{k_1}\ast \dots \ast D_{k_l}}{k_1(k_1+k_2) \dots
   (k_1+ \dots +k_l)}.
\qquad\qquad\mathqed
\end{eqnarray*}}
\hideqed
\end{proof}

Now we compute $\gamma=\gamma\circ I$, where~$I$ is expanded
according to the previous lemma, yielding:
 \allowdisplaybreaks{
\begin{eqnarray*}
    \gamma &=& e + \gamma \circ\bigg\{\sum\limits_{n\in\N^\ast}\,
    \sum_{\substack{k_1, \dots ,k_l\in \N^\ast \\ k_1 + \dots + k_l=n}}\,
    \frac{D_{k_1}\ast \dots \ast D_{k_l}}{k_1(k_1 + k_2)\dots(k_1 +
    \dots + k_l)}\bigg\}
    \\
    &=& e + \sum\limits_{n\in\N^\ast}\,
    \sum_{\substack{k_1,\dots ,k_l\in \N^\ast \\ k_1 + \dots + k_l = n}}\,
    \frac{\gamma \circ D_{k_1}\ast \dots \ast \gamma\circ D_{k_l}}{k_1
    (k_1 + k_2) \dots (k_1 + \dots + k_l)}.
\end{eqnarray*}}
As $D$ preserves the grading, it follows that $\Gamma$ is a left
inverse to the right composition with~$D$.

Similar calculations help to prove that $\Gamma$ is character-valued,
that is, is actually a map from~$\Xi(H)$ to~$G(H)$. Indeed,
let~$\alpha$ be any infinitesimal character in~$\Xi(H)$. Then we have
in $\Char(H)$:
 \allowdisplaybreaks{
\begin{eqnarray*}
         \lefteqn{\Delta (\Gamma (\alpha )) = e \otimes e + \sum_{n >
         0} \sum_{\substack{k_1, \dots, k_l\in\N^\ast \\ k_1 + \dots
         + k_l = n}}\, \frac{\Delta (\alpha_{k_1}\barast \cdots
         \barast \alpha_{k_l})} {k_1(k_1+k_2) \dots (k_1 + \dots +
         k_l)}} \\
         &=& e \otimes e + \sum_{n > 0} \sum_{\substack{k_1, \dots,
         k_l\in\N^\ast \\ k_1 + \dots + k_l = n}}
         \,\sum_{\substack{I\sqcup J = \{k_1,\dots,k_l\} \\
         |I| = m,|J| = p}} \frac{(\alpha_{i_1}\barast \cdots
         \barast \alpha_{i_m}) \otimes(\alpha_{j_1}\barast
         \cdots \barast \alpha_{j_p})} {k_1(k_1 + k_2) \dots
         (k_1 + \dots + k_l)},
\end{eqnarray*}}
where $I=\{i_1,\dots,i_m\},\,J=\{j_1,\dots,j_p\}$ and we have used
that the $\alpha_{i_m}$ are all infinitesimal characters.
Particularly, the assertion we intend to prove, that is $\Delta
(\Gamma(\alpha))= \Gamma(\alpha)\otimes\Gamma(\alpha)$, amounts to
the equality:
 \allowdisplaybreaks{
\begin{eqnarray*}
      \lefteqn{\bigg( e + \sum_{n > 0} \sum_{\substack{k_1, \dots,
      k_l\in\N^\ast \\ k_1 + \dots + k_l =
      n}}\,\frac{\alpha_{k_1}\barast \cdots \barast
      \alpha_{k_l}}{k_1(k_1 + k_2) \dots (k_1 + \dots +
      k_l)}\bigg)^{\otimes 2}} \\
      &=& e \otimes e + \sum_{n > 0} \sum_{\substack{k_1, \dots,
      k_l \in \N^\ast \\ k_1 + \dots + k_l = n}}
      \,\sum_{\substack{I\sqcup J=\{k_1,...,k_l\} \\ |I| = m,|J| = p}}
      \frac{(\alpha_{i_1}\barast \cdots \overline \ast
      \alpha_{i_m})\otimes (\alpha_{j_1}\barast \cdots
      \barast\alpha_{j_p})}{k_1(k_1 + k_2) \dots (k_1 + \dots + k_l)}.
\end{eqnarray*}}
This follows from the identity:
 \allowdisplaybreaks{
\begin{eqnarray}
    \lefteqn{\sum\limits_{I\sqcup J=K}\frac{1}{k_1(k_1 + k_2)
    \dots (k_1 + \dots + k_{p+m})}} \nonumber \\
    &=& \sum_{\substack{i_1,\dots,i_m \in \N^\ast \\ j_1,\dots,j_p \in
        \N^\ast}} \, \frac{1}{i_1(i_1 + i_2)\dots (i_1 + \dots
        + i_m)}\cdot\frac{1}{j_1(j_1 + j_2)\dots (j_1 + \dots + j_p)},
\label{eq:estamos-fritos}
\end{eqnarray}}
where~$K$ runs over all sequences $(k_1,...,k_{p+l})$ obtained by
shuffling the sequences $I$ and $J$.  In turn, the identity follows if
the equation $\Delta (\Gamma (\alpha ))=\Gamma (\alpha )\otimes \Gamma
(\alpha )$ ---that is, $\Gamma (\alpha )\in G(H)$--- holds for a
particular choice of $H$ and $\alpha$ such that the $\alpha_i$ form a
family of algebraically independent generators of a subalgebra of the
convolution algebra $\End(H)$ ---which are therefore also
algebraically independent as elements of~$\Char(H)$.

So, let us consider the particular case $H=T^\ast(X)$ where $H$ is the
graded dual of the enveloping algebra of the FLA over an infinite
alphabet $X$ and $\alpha =D$ is the Dynkin operator.  Then, we already
know that $\Gamma (D)=I$, due to the previous lemma, so that $\Gamma
(D)$ is group-like in $\Char(T^\ast (X))$.  On the other hand, as is
well known, the graded components of the Dynkin operator are
algebraically independent in the convolution algebra
$\End(T^\ast(X))$, and the identity follows.  (For details on the
algebraic independence of the graded components of the Dynkin
operator, consult~\cite{gelfand1995,reutenauer1993}.)

Although we have preferred to give a conceptual proof based on FLAs,
let us mention that identity~\eqref{eq:estamos-fritos} is elementary,
and well known in boson Fock space theory. It also follows e.g. from
the shuffle identity for the product of two iterated
integrals~\cite{shst1993}:
 \allowdisplaybreaks{
\begin{eqnarray*}
    \lefteqn{\sum\limits_{I\sqcup J=K}\int\limits_0^1x_{p+l}^{k_{p+m}-1}
    \dots \int\limits_0^{x_2}x_1^{k_1-1}dx_1\dots dx_{p+m}}\\
    &=& \int\limits_0^1x_m^{i_m-1}\dots
    \int\limits_0^{x_2}x_1^{i_1-1}dx_1\dots dx_m\cdot
    \int\limits_0^1y_p^{j_p-1}\dots
    \int\limits_0^{y_2}y_1^{j_1-1}dy_1\dots dy_p,
\end{eqnarray*}}
which generalizes the integration by parts rule.

\smallskip

To conclude the proof of Theorem~\ref{thm:Gamma} we show that $\Gamma$
is also a right inverse to the composition with~$D$. We contend that,
for any~$h$ in the augmentation ideal of~$H$ and arbitrary $\alpha\in
\Xi(H)$, the following relation holds:
$$
    \alpha (h) = \Gamma(\alpha)^{-1} \ast Y\Gamma(\alpha)\,(h)
          \sepword{or, equivalently,}
    Y \Gamma(\alpha)(h) = \Gamma(\alpha) \ast \alpha\,(h).
$$
Indeed, we have:
 \allowdisplaybreaks{
\begin{eqnarray*}
    Y\Gamma(\alpha)(h) &:=& |h| \sum_{\substack{k_1,\dots,k_l\in \N^\ast
    \\ k_1+ \dots + k_l=|h|}}\, \frac{\alpha_{k_1}\ast \dots \ast
      \alpha_{k_l}}{k_1(k_1+k_2) \dots (k_1+ \dots +k_l)}\,(h)
    \\
    &=& \sum_{\substack{k_1,\dots ,k_l\in \N^\ast \\ k_1+ \dots +
    k_l=|h|}}\, \frac{\alpha_{k_1}\ast \dots \ast
    \alpha_{k_{l-1}}}{k_1(k_1+k_2) \dots (k_1 + \dots + k_{l-1})}\ast
    \alpha_{k_l}\,(h)
    \\
    &=& \Gamma(\alpha) \ast \alpha\,(h).
\end{eqnarray*}}
This together with the fact that $\Gamma(\alpha)\in G(H)$ for
$\alpha \in\Xi(H)$ implies:
\begin{equation*}
    \Gamma(\alpha) \circ D = \Gamma(\alpha)^{-1}\ast Y\Gamma(\alpha) =
    \alpha.
\end{equation*}
Our task is over.\qed

\smallskip

When $H$ is both commutative and cocommutative the convolution product
is commutative and $\gamma\circ D=Y\log\gamma:=\log(\gamma)\circ Y$.
In particular in this case $D=Y\log I$. This was put to good use
in~\cite{bk2000}. Thus clearly $D$, in the general case, is a
noncommutative logarithmic derivative; and the inverse Dynkin
operator~$\Gamma$ a extremely powerful tool. We finally remark that $Y
\ast S$, corresponding, in the free cocommutative case and as an
operator from the tensor algebra to the FLA, to the right-to-left
iteration of bracketings, is another possible form for the
noncommutative logarithmic derivative, leading in particular to
$\gamma \circ (Y\ast S) = Y \gamma \ast \gamma^{-1}$. More generally,
any interpolation of the form $S^a \ast Y \ast S^b$, with $a+b=1$,
yields a notion of noncommutative logarithmic derivative.


\section{Algebraic BWH decomposition of characters}
\label{sect:birkhoffdecomp}

In this section we summarize previous research on Rota--Baxter
operators, relevant for our purpose. Let~$H$ be still graded,
connected and commutative and let again $A$ stand for a commutative
unital algebra. Assume the algebra~$A$ to split directly, $A = A_+
\oplus A_-$, into the subalgebras $A_+,A_-$ with $1_A \in A_+$. We
denote the projectors to~$A_\pm$ by~$R_\pm$, respectively. The pair
$(A,R_-)$ is a special case of a (weight one) Rota--Baxter
algebra~\cite{egk2005} since $R_-$, and similarly $R_+$, satisfies the
relation:
\begin{equation}
    R_-(x)\cdot R_-(y) + R_-(x\cdot y) = R_-\big(R_-(x)\cdot y +
    x\cdot R_-(y)\big), \qquad x,y \in A.
    \label{eq:RBR}
\end{equation}
The reader may verify that integration by parts rule is just the
weight zero Rota--Baxter identity, that is, the second term on left
hand side of~\eqref{eq:RBR} is absent. One easily shows that
$\Lin(H,A)$ with the idempotent operator $\mathcal{R}_-$ defined by
$\mathcal{R}_-(f)=R_- \circ f$, for $f\in \Lin(H,A)$, is a (in general
not commutative) unital Rota--Baxter algebra~\cite{egk2005}.

The subspace $L^{(1)}$ of~$\Lin(H,A)$ made of linear maps that
send the Hopf algebra unit to zero forms a Lie algebra with
$\Xi(A)\subset L^{(1)}$ as a Lie subalgebra. To $L^{(1)}$ does
correspond the group $G_0 = e + L^{(1)}=\exp(L^{(1)})$ of linear
maps sending the Hopf algebra unit to the algebra unit. It
contains the group of characters~$G(A)$ as the subgroup~$\exp(\Xi
(A))$. Due to the characterization of infinitesimal characters as
maps that vanish on the square of the augmentation ideal of~$H$,
both $\mathcal{R}_+(\Xi (A))$ and $\mathcal{R}_-(\Xi (A))$ embed
into~$\Xi(A)$. In particular, both are Lie subalgebras
of~$\Xi(A)$.

The Lie algebra decomposition, $\Xi(A)=\mathcal{R}_+(\Xi (A))\oplus
\mathcal{R}_-(\Xi (A))$ lifts to the group of characters $G(A)$ as
follows. Recall the Baker--Campbell--Hausdorff~(BCH)
formula~\cite{reutenauer1993} for the product of two exponentials,
that we write:
$$
    \exp(x)\exp(y) = \exp\big(x + y + \BCH(x,y)\big).
$$
In~\cite{egk2005,egk2004} the following non-linear map was defined,
whose properties where further explored in~\cite{egm2006}. See
also~\cite{manchon2001}. For $f\in L^{(1)}$, define
$\chi^{\mathcal{R}_-}(f)=\lim_{n \to \infty}
\chi^{\mathcal{R}_-}_n(f)$ where $\chi^{\mathcal{R}_-}_n(f)$ is given
by what we call the BCH recursion:
\allowdisplaybreaks{
\begin{eqnarray}
        \chi^{\mathcal{R}_-}_0(f) &=& f, \qquad
        \chi^{\mathcal{R}_-}_1(f) = f - \BCH\big(\mathcal{R}_-(f),
        \mathcal{R}_+(f)\big),\ \ldots\ ,
        \nonumber \\
        \chi^{\mathcal{R}_-}_{n+1}(f) &=& f - \BCH\Big(\mathcal{R}_-
        \big(\chi^{\mathcal{R}_-}_n(f)\big),
        \mathcal{R}_+\big(\chi^{\mathcal{R}_-}_n(f)\big) \Big).
        \nonumber
\end{eqnarray}}
Then the fixed-point map $\chi^{\mathcal{R}_-}: L^{(1)} \to L^{(1)}$
satisfies:
\begin{equation}
\label{eq:BCHrecursion1}
    \chi^{\mathcal{R}_-}(f) = f -
    \BCH\Big(\mathcal{R}_-\big(\chi^{\mathcal{R}_-}(f)\big),
    \mathcal{R}_+\big(\chi^{\mathcal{R}_-}(f)\big)\Big).
\end{equation}
The superscript ${\mathcal{R}_-}$ is justified by the dependency of
the limit on the Rota--Baxter operator, and the following result.

\begin{lem}
\label{simpleCHI} The map $\chi^{\mathcal{R}_-}$
in~\eqref{eq:BCHrecursion1} solves the simpler recursion:
\begin{equation*}
    \chi^{\mathcal{R}_-}(f) = f + \BCH\Big(-\mathcal{R}_-
    \big(\chi^{\mathcal{R}_-}(f)\big),f\Big), \quad f \in L^{(1)}.
\end{equation*}
\end{lem}

Following \cite{egm2006}, the following factorization theorem holds.

\begin{thm}
\label{thm:bch} For any $f \in L^{(1)}$, we have the unique
factorization:
\begin{equation*}
    \exp(f) = \exp\Big(\mathcal{R}_ -
    \big(\chi^{\mathcal{R}_-}(f)\big)\Big)
     \ast
    \exp\Big(\mathcal{R}_+ \big(\chi^{\mathcal{R}_-}(f)\big)\Big).
\end{equation*}
\end{thm}
Uniqueness of the factorization follows from $R_{-}$ being
idempotent. In the particular case that $f\in \Xi (A)$, the BCH
recursion~\eqref{eq:BCHrecursion1} takes place inside the Lie
algebra~$\Xi(A)$, and the decomposition of $\exp(f)$ holds
therefore inside the group of characters $G(A)$. In particular, it
follows ultimately from Theorem~\ref{thm:bch} that $G(A)$
decomposes as a set as the product of two subgroups:
$$
G(A) = G_-(A)\ast G_+(A), \sepword{where} G_-(A) =
\exp(\mathcal{R}_-(\Xi(A))), \; G_+(A) =\exp(\mathcal{R}_+(\Xi (A))).
$$

\begin{cor} \label{groupfact}
For any $\gamma=\exp(\alpha)\in G(A)$, with $\alpha\in\Xi(A)$, we
have unique $\alpha_{\pm}:=\mathcal{R}_{\pm}
(\chi^{\mathcal{R}_{-}}(\alpha))\in\mathcal{R}_{\pm}(\Xi(A))$, and
unique characters $\gamma_\pm:=\exp(\pm\alpha_{\pm})\in
G_{\pm}(A)$ such that:
\begin{equation}
    \gamma = \gamma_-^{-1} \ast \gamma_+.
\label{eq:BCHbirkhoff}
\end{equation}
\end{cor}

A remark is in order. The
factorization in Theorem~\ref{thm:bch} is true for any (filtration
preserving) linear map $P$ on $\Lin(H,A)$, that is, for
$\chi^{P}(f)$, see~\cite{egm2006}. Uniqueness follows from $P$
being idempotent. The Rota--Baxter property~\eqref{eq:RBR} implies
that both, $G_\pm(A)$ are subgroups. We may reformulate the last
statement about the nature of $G_\pm(A)$ in the next lemma, saying
that the BWH decomposition of Connes and Kreimer, originally found
by a more geometrical method~\cite{ck2000}, is recovered from
Theorem~\ref{thm:bch} by using the Rota--Baxter
relation~\cite{egk2005,Echo,egm2006}:

\begin{lem} \label{ck-Birkhoff}
{}For any $\gamma=\exp(\alpha)$ the unique characters
$\gamma_\pm:=\exp(\pm\alpha_{\pm})\in G_{\pm}(A)$ in the previous
corollary solve the equations:
\begin{equation}
    \gamma_{\pm} = e \pm \mathcal{R}_{\pm}(\gamma_{-} \ast (\gamma - e)).
\label{eq:BogoliubovFormulae}
\end{equation}
\end{lem}
\begin{proof}
There is a sharpened version~\cite{egk2004} of Atkinson's
theorem~\cite{atk1963}, stating that the characters
$\gamma,\gamma_{\pm}$ of~\eqref{eq:BCHbirkhoff} verify
$\gamma_-=e- \mathcal{R}_-(\gamma_{-}\ast(\gamma-e))$ and
$\gamma_+=e- \mathcal{R}_+(\gamma_{+}\ast(\gamma^{-1}-e))$. Now:
$$
    \gamma_{+} \ast (\gamma^{-1}-e) = \gamma_- \ast \gamma \ast
    (\gamma^{-1}-e) = \gamma_-\ast (e - \gamma)
$$
gives~\eqref{eq:BogoliubovFormulae}.
\end{proof}


\section{On renormalization procedures}
\label{sect:reminder}

Prima facie in pQFT, to a Feynman graph~$F$ does correspond by the
Feynman rules a multiple $d$-dimensional momentum space
integral. Each independent loop in a diagram yields one
integration:
\begin{equation}
F \mapsto J_F(p) = \bigg[\int\prod_{l=1}^{|F|}\,d^dk_l \bigg]I_F(p,k).
\label{eq:stone-of-contention}
\end{equation}
Here $|F|$ is the number of loops, $k=(k_1,\ldots,k_{|F|})$ are
the $|F|$ independent internal (loop) momenta and
$p=(p_1,\ldots,p_N)$, with $\sum_{k=1}^N p_k=0$, denote the~$N$
external momenta. In the most relevant kind ($d=4$,
renormalizable with dimensionless couplings) of field theories,
these integrals suffer from ultraviolet (UV) divergencies.
Concretely, under a scale transformation, the integrand behaves as
$$
\bigg[\prod_{l=1}^{|F|}d^d(\lambda k_l)\bigg] I_F(\lambda p,\lambda k)
\sim \lambda^{s(F)},
$$
with~$s(F)$ the superficial UV degree of divergence of the
graph~$F$. Power-counting renormalizable theories are such that
all interaction terms in the Lagrangian are of dimension~$\le d$;
then $s(F)$ is bounded by a number independent of the order of the
graph.  For instance in the $\vf^4_4$~model the superficial UV
degree of divergence of a graph with~$N$ external legs is:
$$
    s(F) = 4 - N.
$$
The Weinberg--Zimmermann theorem says: ``provided all free propagators
have nonzero masses, the integral associated to the Feynman graph~$F$
is absolutely convergent if its superficial UV degree of divergence
and that of each of its one-particle irreducible (1PI) subgraphs is
strictly negative''. The BPHZ momentum space subtraction method is
rooted in this assertion: the idea is to redefine the integrand
$I_F(p,k)$ of a divergent integral by subtracting from it the first
terms of its Taylor expansion in the external momenta~$p$, after these
subtractions have been performed on the integrands corresponding to
the 1PI subgraphs of~$F$ that are renormalization parts; in such a way
the UV degrees of the integral and its subintegrals are lowered until
they become all negative. The combinatorics of those subgraph
subtractions leads to Bogoliubov's recursive $\bar R$-operation and
Zimmermann's forest formula; we assume the reader acquainted with the
former at least~\cite{CaswellK1982,collins1984,smirnov1991,Vasilev04}.

\smallskip

Less straightforward conceptually, but way more practical, is the
DR~method~\cite{h73}.  This belongs to the class of regularization
prescriptions, which parameterize the divergencies appearing in $J_F$
upon introducing non-physical parameters, thereby rendering them
formally finite.  In DR one introduces a complex parameter
$\varepsilon \in \mathbb{C}$ by changing the integral measure, that
is, the space-time dimension, to $\mathrm{D}\in\CC$:
$$
d^dk \xrightarrow{\rm dim\,reg} \mu^{\varepsilon}\,d^\mathrm{D}k,
$$
where $\varepsilon=(d-\mathrm{D})$.  Henceforth always $d=4$.  The
arbitrary parameter $\mu\neq0$ ('t~Hooft's `unit-mass' parameter) is
introduced for dimensional reasons.  Take the~$\vf^4_4$~model: if we
wrote the interaction term simply in the form $g\vf^4/4!$, then the
(naive) dimension of~$g$ would be $[g]=\mu^\eps$.  The redefinition
$\tilde{g}\mu^\eps\vf^4/4!$ of the original vertex in the Lagrangian
includes the mass parameter~$\mu$, introduced to make~$\tilde{g}$
dimensionless.  Now, in any given theory there is a rigid relation
between the numbers of loops and vertices, for each given $N$-point
function.  For instance in the $\vf^4_4$~model, for graphs associated
to the 2-point function the number of vertices is just equal to the
number of loops.  For graphs associated to the 4-point function the
number of vertices is equal to the number of loops plus one, and so an
extra $\tilde{g}\mu^\eps$ factor comes in; but, because we are
correcting the vertex, this extra factor becomes the coupling
constant, and is not expanded in powers of the regularization
parameter~$\eps$; only the expression multiplying it contributes to
the renormalization constant $Z_g$ ---with a pole term independent of
the mass scale~$\mu$.  The outcome is that in practice one computes as
many $\mu^\eps$ factors as loops:
\begin{equation}
\label{mugamma}
    F \longrightarrow J_F^{(\varepsilon,\mu)}(p) =
    \mu^{|F|\varepsilon} \bigg[\int \prod_{l=1}^{|F|} d^\mathrm{D}k_l
    \bigg] \ I_F(p,k).
\end{equation}
See the discussions in~\cite{ck2000,ck2001}
and~\cite[Sections~7~and~8]{greendanger2001} as well. The point is
important from the algebraic viewpoint, since it makes the grading
of Feynman graphs by the number of loops immediately relevant to
the renormalization process.

The next step in~DR consists of a specific subtraction rule of
those $\varepsilon$-parameterized expressions which allows to take
the limit $\varepsilon \downarrow 0$. Now, Connes and Kreimer's
BWH decomposition of Feynman rules~\cite{ck2000} is
extraordinarily well adapted to~DR in~pQFT. In the Connes--Kreimer
paradigm, any renormalizable quantum field theory gives rise to a
Hopf algebra~$H$ of Feynman graphs, polynomially generated by 1PI
Feynman graphs and graded by graph loop number. The coproduct
operation of~$H$ mirrors the combinatorics of the subgraphs.
Looking back to~\eqref{eq:stone-of-contention}, the unrenormalized
Feynman integral does define a character because:
$$
    I_{F_1 \cup F_2}(p_1,p_2,k_1,k_2) =
    I_{F_1}(p_1,k_1)I_{F_2}(p_2,k_2)
$$
for disjoint graphs $F_1,F_2$.  On the Hopf algebra~$H$ the
Feynman rules single out a distinguished character~$\gamma$ with
values in a suitable target algebra of regularized amplitudes.
Precisely, Connes and Kreimer establish the above decomposition
$G(A)=G_{-}(A)\ast G_{+}(A)$, for~$A$ the algebra of
amplitude-valued Laurent series in the dimension
defect~$\varepsilon$, using the MS scheme in DR on momentum space.
The characters $\gamma_{\pm}$ in the
decomposition~\eqref{eq:BCHbirkhoff} solve Bogoliubov's
renormalization formulae ---see Corollary~\ref{BogoliubovMap}
below--- and may be called the renormalized and counterterm
character, respectively.  The sought after result is given by the
`positive' part in~\eqref{eq:BCHbirkhoff}.  To wit,
$\gamma_+^{(\varepsilon,\mu)}= \gamma_-^{(\varepsilon,\mu)}\ast
\gamma^{(\varepsilon,\mu)}$, and the limit $\gamma_+^{(\varepsilon
\downarrow 0,\mu)}$ exists, giving the renormalized Feynman
amplitude.  In what follows, when dealing with dimensionally
regularized characters we drop the superscript~$\varepsilon$.  We
also forget about the other involved variables, that do not enter
our considerations, and just write
$\CC[[\varepsilon,\varepsilon^{-1}]$ for~$A$.  Thus $R_-$ will be
the projection onto the subalgebra $A_-:= \varepsilon^{-1}
\mathbb{C}[\varepsilon^{-1}]$.  In summary, $A$ encodes the
Feynman rules within the chosen regularization procedure, whereas
the splitting of~$A$, hence the projector~$R_-$, reflects a
renormalization scheme within that choice.

\begin{cor} \label{BogoliubovMap}
The map $\bar{\gamma}:=\gamma_{-} \ast (\gamma - e) = \gamma_+ -
\gamma_-$ in~\eqref{eq:BogoliubovFormulae} gives Bogoliubov's
preparation map~$\bar R$.
\end{cor}

\noindent Indeed with the Connes--Kreimer definition of the Hopf
algebra of Feynman graphs, equations~\eqref{eq:BogoliubovFormulae}
coincide with Bogoliubov's recursive formulas for the counterterm
and renormalized parts.

Recalling the remark after
Corollary~\ref{groupfact} we may characterize the Rota--Baxter
structure on $A$ as follows. Theorem~\ref{thm:bch} implies that
regularized, i.e. $A$-valued, Feynman rules on $H$ factorize
uniquely upon the choice of an idempotent linear map on $A$
respectively $\Lin(H,A)$. The extra demand for a Rota--Baxter
structure on $A$, and hence on $\Lin(H,A)$ essentially implies the
particular recursive nature of the process of perturbative
renormalization as encoded in Bogoliubov's preparation map~$\bar
R$ respectively the equations~(\ref{eq:BogoliubovFormulae}).


\section{Locality and the Connes--Kreimer beta function}
\label{sect:locality}

The results in Sections~\ref{sect:HopfCharacters}
to~\ref{sect:birkhoffdecomp} apply to any graded connected commutative
bialgebra $H$ and any commutative unital algebra $A$ with a direct
splitting. In this section we restrict our consideration to the class
of Hopf algebra characters possessing a \textit{locality} property,
with~$H$ as before. This will correspond to the example given by the
Feynman rules for a renormalizable pQFT in DR, using the framework of
the~MS and the~$\overline{\rm MS}$~\cite[Section~5.11.2]{collins1984}
schemes. There locality comes from the dependency of DR on the
arbitrary `mass parameter'~$\mu$. It is handy to provisionally fix the
value of~$\mu$ at some convenient reference point~$\mu_0$. The
difference between both schemes boils down to:
\begin{equation}
    \mu_0 = {\overline\mu}_0\frac{e^{\gamma_E/2}}{2\sqrt\pi},
\label{eq:tapa-del-perol}
\end{equation}
with $\mu_0,{\overline\mu}_0$ respectively the MS, $\overline{\rm MS}$
values and $\gamma_E$ the Euler--Mascheroni constant. Our aim is to
recover by our methods the results in~\cite{ck2001,cm22004}; the latter
constitute a stronger, algebraic version of a famous theorem
by~'t~Hooft~\cite{h73}. Place of pride corresponds to
the Connes--Kreimer abstract beta function. This is a conceptually very
powerful beast, giving rise to the ordinary beta function through the
(tangent morphism to) the morphism (of unipotent groups) from~$G(\CC)$
to the group of transformations of the coupling
constants~\cite{ck2001}.

Any $f \in \Lin(H,A)$ is now of the form:
$$
    f(h) = \sum_{k=-U}^\infty f_{:k}(h)\varepsilon^k =: f(h;\varepsilon)
$$
for $h \in H$. Here every $f_{:k}\in \Lin(H,\CC )$, the dual of $H$;
and the last notation is used when we want to emphasize the dependency
on the~DR parameter~$\varepsilon$. If~$h$ is a $|h|$-loop 1PI Feynman
graph, a general theorem~\cite{smirnov1991} in DR says that $U = |h|$
at non-exceptional momenta.

We define on the group of $A$-valued characters $G(A)$ a one-parameter
action of~$\CC^* \owns t$ given for~$h$ homogeneous in~$H$ by:
\begin{equation}
    \psi^t(h;\varepsilon) := t^{\varepsilon|h|}\psi(h;\varepsilon).
\label{eq:pesar-de-los-pesares}
\end{equation}
Physically this amounts to neatly replacing the $\mu_0^{\eps|h|}$
factor present in each dimensionally regularized Feynman
diagram~\eqref{mugamma} by $(\mu_0 t)^{\eps|h|}$; that is, the mass
scale is changed from~$\mu_0$ to~$t\mu_0$ ---or
from~${\overline\mu}_0$ to~$t{\overline\mu}_0$, as the case might be.
As~$\eps$ is a complex variable, there is no harm in taking~$t$
complex, too.

It is clear that $\psi^t$ in~\eqref{eq:pesar-de-los-pesares} is
still a character, and that $(\psi_1\ast
\psi_2)^t=\psi_1^t\ast\psi_2^t$. This last property also holds if
$\psi_1,\psi_2$ in~\eqref{eq:pesar-de-los-pesares} belong more
generally to~$\Lin(H,A)$. Besides, for future use we store:
\begin{equation}
    t\frac{\partial}{\partial t}\psi^t =
    \varepsilon |h|\psi^t(h;\varepsilon)
    = \varepsilon\,Y\psi^t\ \quad {\rm{such\ that}} \quad
    t\frac{\partial}{\partial t}\Big|_{t=1}\,\psi^t = \varepsilon\,Y\psi.
\label{eq:hour-of-reckoning}
\end{equation}
For any~$t$ and any homogeneous~$h \in H$ we have
$t^{\varepsilon|h|}\in\mathcal{R}_+(A)=A_+:=
\mathbb{C}[[\varepsilon]]$, so that the one-parameter action on~$G(A)$
restricts to a one-parameter action on the group~$G_+(A)$~:
$$
    \psi \in G_+(A) \mapsto \psi^t \in G_+(A).
$$
We now have for the regularized, but unrenormalized character
$\gamma^t \in G(A)$ a BWH decomposition:
$$
    \gamma^t = {(\gamma^t)}_-^{-1} \ast {(\gamma^t)}_+.
$$
Notice that we write instead $(\gamma_-)^t$ and~$(\gamma_+)^t$ for the
image of~$\gamma_-$ and~$\gamma_+$ under the one-parameter group
action. The locality property simply states that for the counterterm
the following holds.

\begin{thm}
\label{loctheom}
Let $\gamma$ be a dimensionally regularized Feynman rule
character. The counterterm character in the BWH decomposition
$\gamma^{t}= (\gamma^{t})^{-1}_-\ast(\gamma^{t})_+$ satisfies:
\begin{equation}
\label{locality}
    t\,\frac{\partial{(\gamma^t)}_-}{\partial t} = 0 \sepword{or\
    ${(\gamma^t)}_-$ is equal to~$\gamma_-$, i.e. independent of~$t$.}
\end{equation}
We say the $A$-valued characters $\gamma \in G(A)$ with this property
are \textit{local} characters.
\end{thm}

The physical reason for this is that the counterterms can be taken
mass-independent; this goes back at least to~\cite{ArcheoCollins}. For
this fact, and more details on the physical significance of the
locality property in pQFT, we refer the reader
to~\cite{bergkrei2005,collins1984,ck2000,cm22004}.

In the sequel, albeit mustering Dynkin technology, we partly
follow the plan in reference~\cite{manchon2001}. Denote by $G^{\rm
loc}(A)$ the subset of local elements of~$G(A)$ and $G^{\rm
loc}_-(A)$ the subset of elements in~$G^{\rm loc}(A)$ for which
$\psi(h) \in A_-$ when~$h$ has no scalar part.

\begin{prop}\label{prop:BWHloc}
The set $G^{\rm loc}(A)$ decomposes into the product $G^{\rm
loc}_-(A)\ast G_+(A)$.
\end{prop}

\begin{proof}
Notice first that $G_+(A)$ embeds in $G^{\rm loc}(A)$, since
$\psi^t\in G_+(A)$ for any $\psi\in G_+(A)$. Next, if~$\phi$ is
local and $\rho \in G_+(A)$, then $\phi \ast \rho$ is local.
Indeed, we have:
$$
    \phi^t \ast \rho^t = {(\phi^t)}_-^{-1} \ast {(\phi^t)}_+ \ast
    \rho^t,
$$
with polar part the one of $\phi^t$, which is constant since $\phi$ is
local. Let~$\phi$ still be local. Then $\phi \ast \phi_+^{-1} =
\phi_-^{-1}$, and the proposition follows if we can show $\phi_-^{-1}
\in G^{\rm loc}_-(A)$. Now, if~$\phi$ is local, we have:
$$
    (\phi_-^{-1})^t \ast (\phi_+)^t = \phi^t = {(\phi^t)}_-^{-1} \ast
    {(\phi^t)}_+;
$$
so that the BWH decomposition of~$(\phi_-^{-1})^t$ is given by:
\begin{equation}
    (\phi_-^{-1})^t = {(\phi^t)}_-^{-1} \ast {(\phi^t)}_+ \ast
    ((\phi_+)^t)^{-1},
\label{eq:RG1}
\end{equation}
with polar part ${(\phi^t)}_-$, the one of~$\phi$, constant and
equal to~$\phi_-$.
\end{proof}

Now we wish to complete the study of locality by understanding how the
decomposition of $G^{\rm loc}(A)$, which is a refinement of the
decomposition $G(A)=G_-(A)\ast G_+(A)$, is reflected at the Lie
algebra level.  More precisely, we would like to know if there is a
subspace of~$\mathcal{R}_-(\Xi(A))$ naturally associated to~$G_-^{\rm
loc}(A)$.  The answer (quite surprising at first sight) is simple and
well known to physicists: beta functions are enough.  As shown below,
an excellent tool to understand this is the Dynkin operator pair.  Let
now $\beta\in\Xi(\CC)$ be a \textit{scalar}-valued infinitesimal
character.  Notice that $\beta/\varepsilon$ can be regarded as an
element of~$\mathcal{R}_-(\Xi(A))$.

\begin{prop}\label{prop:GammaBeta}
With $\Gamma$ as defined in Eq.~(\ref{eq:madre-del-cordero}) of
Theorem~\ref{thm:Gamma}, we find:
\begin{eqnarray*}
    \psi_\beta &:=& \Gamma(\beta/\varepsilon) \in G^{\rm loc}_-(A).
\end{eqnarray*}
\end{prop}

\begin{proof}
{}From Eq.~\eqref{eq:madre-del-cordero} it is clear that~:
\begin{equation}
\label{eq:GammaBeta}
     \psi_\beta = \Gamma(\beta/\varepsilon) = \sum\limits_n
     \bigg(\,\sum\limits_{k_1, \dots,k_n\in \N^\ast}
     \frac{\beta_{k_1} \ast \dots\ast \beta_{k_n}} {k_1(k_1 + k_2)
     \dots (k_1 + \dots + k_n)}\bigg) \frac{1}{\varepsilon^{n}}
\end{equation}
implying $\psi_\beta \in G_-(A)$. Next we observe $\psi_\beta^t =
\Gamma(\beta^t/\varepsilon)$, which follows simply
from~\eqref{eq:GammaBeta} and, on use
of~\eqref{eq:hour-of-reckoning}:
$$
   (\psi_\beta^t)^{-1} \ast t\frac{\partial}{\partial t}\psi_\beta^t
   = \varepsilon {(\psi_\beta^t)}^{-1} \ast Y \psi_\beta^t
   = \varepsilon \psi_\beta^t\circ D
   = \varepsilon \Gamma(\beta^t/\varepsilon) \circ D
   = \beta^t.
$$
Now, the BWH decomposition $\psi_\beta^t = (\psi_\beta^t)^{-1}_-
\ast(\psi_\beta^t)_+$ is such that~:
 \allowdisplaybreaks{
\begin{eqnarray*}
    (\psi_\beta^t)^{-1} \ast t\frac{\partial}{\partial t} \psi_\beta^t
    &=& (\psi_\beta^t)^{-1} \ast
         t\frac{\partial}{\partial t} {(\psi_\beta^t)}^{-1}_- \ast
        {(\psi_\beta^t)}_+ + (\psi_\beta^t)^{-1} \ast
        {(\psi_\beta^t)}^{-1}_- \ast
        t\frac{\partial}{\partial t}{(\psi_\beta^t)}_+ \\
    &=& {(\psi_\beta^t)}_+^{-1} \ast {(\psi_\beta^t)}_- \ast
        t\frac{\partial}{\partial t}{(\psi_\beta^t)}^{-1}_-
        \ast{(\psi_\beta^t)}_+ + {(\psi_\beta^t)}^{-1}_+
        \ast t\frac{\partial}{\partial t}{(\psi_\beta^t)}_+;
\end{eqnarray*}}
hence we find:
 \allowdisplaybreaks{
\begin{eqnarray*}
    {(\psi_\beta^t)}_+ \ast \beta^t \ast {(\psi_\beta^t)}_+^{-1} &=&
    {(\psi_\beta^t)}_- \ast t\frac{\partial}{\partial
    t}{(\psi_\beta^t)}^{-1}_- + t\frac{\partial}{\partial
    t}{(\psi_\beta^t)}_+ \ast {(\psi_\beta^t)}^{-1}_+.
\end{eqnarray*}}
Using $\psi_\beta \in G_-(A)$ and that $\beta^t$ takes values
in~$A_+$, we find by applying the projector~$\mathcal{R}_-$ on
both sides of the last equation~:
\begin{equation*}
    \mathcal{R}_-\Big( (\psi_\beta^t)_+ \ast \beta^t \ast
    (\psi_\beta^t)_+^{-1} \Big) = 0 = (\psi_\beta^t)_- \ast
    t\frac{\partial}{\partial t} (\psi_\beta^t)^{-1}_- = -
    t\frac{\partial}{\partial t}(\psi_\beta^t)_- \ast
    (\psi_\beta^t)^{-1}_-
\end{equation*}
implying that ${(\psi_\beta^t)}_-$ is independent of~$t$; thus
$\Gamma(\beta/\varepsilon)$ is a local character.
\end{proof}

The last proposition is suggestive of the fact that local $A$-valued
characters are essentially determined by \textit{ordinary}
(scalar-valued) infinitesimal characters (beta functions). This is
indeed the case. Before proving the main result of this section, we
remark that, for any $\phi\in G^{\rm loc}(A)$, we can write~:
\begin{equation}
    \phi^t = \phi_-^{-1} \ast (\phi^t)_+ = \phi \ast h_{\phi}^t,
\label{eq:localBHW}
\end{equation}
where $h_{\phi}^t := (\phi_+)^{-1} \ast (\phi^t)_+ \in G_+(A)$.
Also, for $\phi\in G_-^{\rm loc}(A)$ we denote $\phi_{:-1}$
by~$\Res\phi$.

\begin{thm}
\label{thm:residue} The map $\phi \mapsto \varepsilon(\phi\circ
D)$, with~$D$ the Dynkin operator, sends $G^{\rm loc}(A)$ to
$\Xi(A_+)$ and $G^{\rm loc}_-(A)$ to $\Xi(\CC)$; explicitly, in
the second case:
$$
    G^{\rm loc}_-(A) \ni \phi \mapsto
    \varepsilon(\phi \circ D) = Y\Res\phi \in \Xi(\CC).
$$
\end{thm}

\begin{proof}
First, recall from Corollary~\ref{cor:DynkinD} that $D$ sends
characters to infinitesimal characters and that by
Proposition~\ref{prop:BWHloc} any local $\phi \in G^{\rm loc}(A)$
decomposes as $\phi^a \ast \phi^b$ with $\phi^a \in G_-^{\rm
loc}(A), \phi^b\in G_+(A)$. Therefore, we see that:
$$
    \phi \circ D = \phi^{-1}\ast Y\phi
                 = (\phi^b)^{-1}\ast (\phi^a)^{-1}\ast Y\phi^a\ast \phi^b
                   + (\phi^b)^{-1}\ast Y\phi^b;
$$
since $(\phi^b)^{-1}$ and $Y\phi^b$ belong to $\Lin(H,A_+)$, the
theorem follows if we can prove that:
$$
    \epsilon(\phi\circ D) = \epsilon(\phi^{-1}\ast Y\phi)\in \Xi(\CC )
$$
when $\phi\in G_-^{\rm loc}(A)$. Assume the latter is the case and
recall the decomposition $\phi^t = \phi \ast h_{\phi}^t$
in~\eqref{eq:localBHW}, with now simply $h_{\phi}^t=(\phi^t)_+$. So
that on the one hand~\eqref{eq:hour-of-reckoning} implies~:
$$
      t\frac{\partial}{\partial t}\Big|_{t=1}h^t_\phi
      = \phi^{-1} \ast t\frac{\partial}{\partial t}\Big|_{t=1}\phi^t
      = \phi^{-1} \ast \varepsilon\,Y\phi
      = \varepsilon(\phi \circ D).
$$
On the other hand, observe that by using the Bogoliubov
formula~\eqref{eq:BogoliubovFormulae} one finds:
 \allowdisplaybreaks{
\begin{eqnarray}
     t\frac{\partial}{\partial t}\Big|_{t=1}h^t_\phi &=&
     t\frac{\partial}{\partial t}\Big|_{t=1}(\phi^t)_+ =
     t\frac{\partial}{\partial t}\Big|_{t=1} \mathcal{R}_+\big(\phi_-
     \ast (\phi^t - e)\big)\\
     &=& t\frac{\partial}{\partial t}\Big|_{t=1}
     \mathcal{R}_+\big(\phi^{-1} \ast \phi^t - \phi^{-1}\big) =
     t\frac{\partial}{\partial t}\Big|_{t=1}
     \mathcal{R}_+\big(\phi^{-1} \ast t^{\varepsilon|\cdot|}\phi \big)
     \nonumber \\
    &=& \mathcal{R}_+\big(\varepsilon\,Y\phi) = Y\Res\phi.
\end{eqnarray}}
In the step before the last we took into account that $Y(1_H)=0$
and that $\phi_- \in G_-^{\rm loc}(A)$ which implies for $h \in
H^+$:
$$
    \phi^{-1} \ast t^{\varepsilon|\cdot|}\phi(h) = \phi^{-1}(h) +
    t^{\varepsilon|h|}\phi(h) + \phi^{-1}(\overline{h}^{(1)})
    t^{\varepsilon|\overline{h}^{(2)}|}\phi(\overline{h}^{(2)}),
$$
where $h^{(1)}\otimes h^{(2)}=h\otimes 1+1\otimes h+
\overline{h}^{(1)}\otimes\overline{h}^{(2)}$. Here $\phi^{-1}(h)\in
A_-$ and $\phi^{-1}(\overline{h}^{(1)})
t^{\varepsilon|\overline{h}^{(2)}|}\phi(\overline{h}^{(2)})$ are both
mapped to zero by $t\frac{\partial}{\partial
t}\Big|_{t=1}\mathcal{R}_+$.
\end{proof}

A glance back to Theorem~\ref{thm:Gamma} and
Proposition~\ref{prop:GammaBeta} allows us to conclude that for
$\phi\in G^{\rm loc}_-(A)$ one has:
\begin{equation}
    \Gamma\bigg(\frac{Y\Res\phi}{\varepsilon}\bigg) = \phi,
    \label{eq:dies-illa}
\end{equation}
so indeed the polar part of a local character $\phi$ can be
retrieved from its beta function, $\beta(\phi):= Y\Res\phi \in
\Xi(\mathbb{C})$, by the universal
formula~\eqref{eq:madre-del-cordero}.

Equation~\eqref{eq:dies-illa} is equivalent to the `scattering
type formula' of~\cite{ck2001}; we can refer to~\cite{manchon2001}
for this. Now it is an easy task to recover the other results
of~\cite{ck2001,cm22004}. We ought to remark that, for~$\gamma$ a
general local character, one defines $\Res\gamma=-\Res\gamma_-$
---see in this respect formulae~\cite[Equation~(11)]{ck2001}
or~\cite[Equation~(2.111)]{cm22004}.

\begin{thm}
\label{thm:prueba-de-fuego} For the renormalized character
$\gamma_{\rm ren}(t):= (\gamma^t)_+(\eps=0)$ it holds:
\begin{equation}
\label{eq:Trojan-gift}
    t\frac{\partial}{\partial t} \gamma_{\rm ren}(t) = (Y \Res\gamma)
    \ast \gamma_{\rm ren}(t),
\end{equation}
the abstract RG~equation.
\end{thm}

\begin{proof}
First, in the proof of Theorem~7.2 we saw already that~$D$ verifies a
cocycle condition~\cite{em2006}: for $\phi,\psi\in G(A)$:
\begin{equation*}
    (\phi \ast \psi) \circ D = \psi^{-1} \ast (\phi \circ D) \ast \psi
    + \psi \circ D.
\end{equation*}
This together with Theorem~\ref{thm:residue} implies for $\phi\in
G_-^{\rm loc}(A)$ that $\Res \phi =-\Res \phi^{-1}$. Indeed, this
follows by taking the residue~$\Res$ on both sides of the equation:
\begin{equation*}
    0 = (\phi^{-1} \ast \phi) \circ D
      = \phi^{-1} \ast (\phi^{-1} \circ D) \ast \phi + \phi \circ D
      = \phi^{-1} \ast \frac{\Res\phi^{-1}}{\varepsilon} \ast \phi +
        \frac{\Res\phi} {\varepsilon}.
\end{equation*}
Now, let $\gamma \in G^{\rm loc}(A)$ with BWH decomposition $\gamma^t
= \gamma^{-1}_- \ast {(\gamma^t)}_+$. Recall that ${(\gamma^t)}_+ =
\mathcal{R}_+\big(\gamma_- \ast t^{\varepsilon|\cdot|}\gamma \big)$
maps~$H^+$ into~$A_+\otimes\mathbb{C}[[\log(t)]]$ such that:
$$
    t\frac{\partial{(\gamma^t)}_+}{\partial t}(0) =
    t\frac{\partial}{\partial t}{\gamma}_{\rm ren}.
$$
As $\gamma_- \in G_-^{\rm loc}(A)$, we then find:
 \allowdisplaybreaks{
\begin{eqnarray*}
    t\frac{\partial(\gamma^t)_+}{\partial t}
    &=& \gamma_- \ast t\frac{\partial}{\partial t}\gamma^t
     = \gamma_- \ast \varepsilon Y \gamma^t
     = \gamma_- \ast \varepsilon Y(\gamma^{-1}_- \ast (\gamma^t)_+)
    \\
    &=& \gamma_- \ast \varepsilon Y(\gamma^{-1}_-) \ast {(\gamma^t)}_+
        + \varepsilon Y{(\gamma^t)}_+ = \varepsilon (\gamma^{-1}_-\circ D)
        \ast {(\gamma^t)}_+ + \varepsilon Y{(\gamma^t)}_+ \\
        &=& (Y\Res\gamma^{-1}_-) \ast {(\gamma^t)}_+ + \varepsilon
        Y{(\gamma^t)}_+ = - (Y\Res\gamma_-) \ast {(\gamma^t)}_+ +
        \varepsilon Y(\gamma^t)_+ \\
    &=& (Y\Res\gamma) \ast {(\gamma^t)}_+ + \varepsilon Y(\gamma^t)_+.
\end{eqnarray*}}
Therefore both sides have a limit as~$\varepsilon\downarrow0$,
yielding the sought after RG equation~\eqref{eq:Trojan-gift}.
\end{proof}

Equation~\eqref{eq:Trojan-gift} is solved using the beta function
$\beta(\gamma):= Y\Res\gamma \in \Xi(\mathbb{C})$:
$$
    \gamma_{\rm ren}(t) = \exp(\ln(t)\beta(\gamma)) \ast
    \gamma_{\rm ren}(1).
$$
The last statement and equation~\eqref{eq:RG1} tell us that:
$$
    \lim_{\varepsilon \to 0} \gamma_{-} \ast (\gamma_-^{-1})^t =
    \lim_{\varepsilon \to 0} (\gamma^{t})_+ \ast ((\gamma_+)^t)^{-1} =
    \gamma_{\rm ren}(t) \ast \gamma_{\rm ren}^{-1}(1) =
    \exp(\ln(t)\beta(\gamma)).
$$
The scalar-valued characters
$$
\Omega_t(\gamma):= \exp(\ln(t)\beta(\gamma)) \in G(\mathbb{C})
$$
obviously form a one-parameter subgroup in~$G(A)$: $\Omega_{t_1}
(\gamma)\ast\Omega_{t_2}(\gamma)=\Omega_{t_1t_2}(\gamma)$, generated
by the beta function and controlling the flow of the renormalized
Feynman rule character with respect to the mass scale.


\section{Through the prism of other renormalization schemes I}
\label{sect:DSlocality}

We plan now to prospect the usefulness of our approach in other
schemes of renormalization. Doubtless DR provides the cleanest
formulation of locality in the BWH decomposition for
renormalization. However, it is physically clear that in any
scheme one has still to parameterize the arbitrariness in
separating divergent from finite parts; and that the physical
irrelevance of the choices made still gives rise to the
RG~equation. On the mathematical side, it is worth to recall that
the algebraic BWH decomposition of
Section~\ref{sect:birkhoffdecomp} is not necessarily linked to
loops on the Riemann sphere. It is thus legitimate to ponder the
question in schemes other than those based on~DR. We plan to
exemplify with the BPHZ~scheme in the next section, but, before
dwelling on that, let us briefly indicate other pieces of
evidence.

A first one concerns old research. In the early seventies, Callan
set out to prove that broken scale invariance~\cite{callan1970} is
all that renormalization was about. He was eventually able to give
a treatment of the RG, and proofs of renormalizability of field
theories based on the former, by relying entirely in the BPHZ
formalism. To derive the beta function, he just set up
RG~equations by studying the dependency of the $N$-point functions
on the renormalized mass. See in this
respect~\cite{by1974,Blue1976}. In a renormalization method
without regularization, information about the~RG must be stored
somehow in the \textit{renormalized} quantities. Concretely, as
hinted at by our last theorem, one finds it in the scaling
properties of the renormalized integral. This was noted in the
context of Epstein--Glaser renormalization in~\cite{jmgb2003}. In
DR this shows in the RG equation~\eqref{eq:Trojan-gift}.

A second piece of evidence is furnished by more recent work by
Kreimer and
collaborators~\cite{bergkrei2005,bk2001,kreimer2006,kreiyea2006}.
Indeed, Kreimer has long argued that locality (and
renormalizability) is determined by the Hochschild cohomology of
renormalization Hopf algebras.  This cohomology is trivial in
degree greater than one.  The coproduct on~$H$ can be written
recursively in terms of closed Hochschild 1-cochains.  Those are
grafting maps indexed by primitive 1PI diagrams, that govern the
structure of Feynman graphs and give rise through the Feynman
rules to integral Dyson--Schwinger equations. Here is not the
place for details and we refer the reader to~
\cite{bergkrei2005,bk2001,kreimer2005,kreimer2006,kreiyea2006},
and especially Kreimer's recent review~\cite{kreimer2006f}.  The
interaction between the Hopf algebra of characteristic functions
of~$H$ of this paper and the Hochschild 1-cocycles on~$H$ is a
promising field of interest.

In the indicated references the Dynkin operator~$D$ (and its close
cousins $S \ast Y^n$) appears, again defining the residue, in
renormalization schemes without regularization. There Green's
functions, $\Sigma=\Sigma(g,p)$, are defined in terms of their
(combinatorial) Dyson--Schwinger equations using the Hochschild
1-cocycles; $g,p$ denote the coupling constant and external momenta,
respectively. Those Green's functions are expanded as power series
in~$g$:
$$
    \Sigma = 1 + \sum_{k>0}\phi(c_k)g^k,
$$
for Feynman rules $\phi \in G(\mathbb{C})$ and with order by order
divergent Feynman amplitudes $\phi(c_k)$ as coefficients. Here the
$c_k$'s are particular linear combinations of graphs of loop order~$k$
in~$H$~\cite{kreimer2005}. Renormalization of~$\Sigma$ is achieved by
using a renormalized character $\phi_{\rm ren}\in G(\mathbb{C})$
defined by subtraction at a specific point $p^2=\lambda^2$
---corresponding to Taylor expansion up to zeroth order.
Here~$\lambda$ plays the role of the mass-parameter~$\mu$.
Locality is automatically fulfilled in this approach. The
renormalized Green's functions $\Sigma_{\rm ren}=\Sigma_{\rm
ren}(g,p,\lambda)$ can be developed in terms of the parameter
$L:=\log(p^2/\lambda^2)$, hence $\Sigma_{\rm ren} = 1 +
\sum_{k>0}\alpha_k(g)L^k$, with $\alpha_1(g) \in
\Xi(\CC)$~\cite{kreiyea2006}. Following the above references and
adapting partially to our notation, the residue is found to be:
$$
    \Xi(\CC) \ni \sigma_1 := \frac{\partial}{\partial L}(\phi_{\rm
    ren} \circ D) \bigg|_{L=0} = \alpha_1(g).
$$
In~\cite{kreiyea2006} Kreimer and Yeats outline how to derive
$\alpha_k(g)$, $k>1$ recursively from $\alpha_1(g)$.  This confirms
that, in a deep sense, the beta function \textit{is} composition with
the Dynkin operator.


\section{Through the prism of other renormalization schemes II}
\label{sect:BPHZlocality}

Let us now explore the classical BPHZ scheme in the context of the
former sections. With~$I_F$ the integrand
of~\eqref{eq:stone-of-contention} corresponding to the graph $F$,
let us write for the Taylor subtraction employed in BPHZ
renormalization:
\begin{equation*}
    I_F(p,k) \mapsto I_F(p,k) - t^{s(F)}_pI_F(p,k) := I_F(p,k) -
    \sum_{|\alpha|\le s(F)}\,\frac{p^{\alpha}}{\alpha!}
    \partial_{\alpha}I_F(0,k).
\end{equation*}
We borrowed here the standard multi-index notation
$$
    \alpha = \{\alpha_1,\dots,\alpha_n\}\in\N^n,
    \quad
    |\alpha|:= \sum_{i=1}^n\alpha_i,
    \quad
    \alpha!=\prod_{i=1}^n\alpha_i!\,;
$$
each $\alpha_i$ takes values between~$0$ and~$3$, say. We are
assuming that only massive particles are present, otherwise the
subtraction at zero momentum gives rise to infrared divergences;
the expression of~$t^{s(F)}_pI_F$ in practice simplifies because
of Lorentz invariance.

Notice that the integral~$J_F(p)$
in~\eqref{eq:stone-of-contention} \textit{does} originally have a
meaning: it is a well defined functional on the linear subspace
$\cS_{s(F)}(\R^{4N})$ of Schwartz functions~$\phi$ on the external
momenta, whose first moments $\int p^\alpha\phi(p)\,d^{4N}p$ up to
order~$|\alpha|\le s(F)$ happen to vanish. The ``divergent'' loop
integrals inside~$J_F(p)$ become harmless when coupled exclusively
with Schwartz functions of this type. The Taylor `jet' projector
map $t^l_p$ subtracts polynomials, that are functionals of the
same type, in such a way that the result (eventually) becomes
moreover a tempered distribution.

The first question is whether we have a Rota--Baxter algebra in the
BPHZ context.  Actually, we do have the Rota--Baxter property
for~$t^l_p$.  Indeed, the following is obtained by a simple
calculation from the definitions.

\begin{prop}
Let $I_{F_i},\,i=1,2$ have associated degrees of divergence
$l_i,\,i=1,2$. Then
\begin{equation*}
    t^{l_1}_{p_1}\big(I_{F_1}\big)\,
    t^{l_2}_{p_2}\big(I_{F_2}\big) =
    t^{l_1+l_2}_{p_1,p_2}\big(I_{F_1}\,t^{l_2}_{p_2}(I_{F_2})\big)
    +
    t^{l_1+l_2}_{p_1,p_2}\big(t^{l_1}_{p_1}(I_{F_1})\,I_{F_2}\big)
    -
    t^{l_1+l_2}_{p_1,p_2}\big(I_{F_1}I_{F_2}\big).
\end{equation*}
\end{prop}

We leave the verification of this to the care of the reader. In
general, if $U$ is a multiplicative semigroup, a family of linear
operators $R_u,\,u\in U$ on the algebra~$A$ is called a Rota--Baxter
family if for any $u,v\in U$ and $a, b\in A$, we have
\begin{equation*}
    R_u(a)R_v(b) = R_{uv}(aR_v(b)) + R_{uv}(R_u(a)b) - R_{uv}(ab),
    \quad \sepword{for all} a,b \in A.
\end{equation*}
Thus the $l$-jets define a Rota--Baxter family. Now, a Rota--Baxter
family is almost the same thing as a Rota--Baxter operator.

\begin{prop} \label{pp:RBF}
Let $\mathcal{A}=A[U]$ be the semigroup algebra associated to~$A$. Let
$R_u: A \to A,\,u\in U$ be a Rota--Baxter family. Define
$$
    R: \mathcal{A} \to \mathcal{A},
    \sepword{by}
    R\Big(\sum_u a_u u\Big) :=\sum_u R_u(a_u) u.
$$
Then $R$ is a Rota--Baxter operator on~$\mathcal{A}$ such that
$R(au)=a'u$ with~$a'$ in~$A$. Conversely, if $R:\mathcal{A}\to
\mathcal{A}$ is a Rota--Baxter operator such that $R(au)=a'u$
with~$a'$ in~$A$, then we obtain a Rota--Baxter family $R_u,u\in
U$, by defining $R_u(a)=a'$ where $R(a\,u)=a'\,u$.
\end{prop}

The proof is immediate\footnote{We thank L.~Guo for suggesting the
notion of Rota--Baxter family}. On the strength of the previous
result, we may refer to a Rota--Baxter family as a Rota--Baxter
operator. Now, pQFT in practice furnishes an even more radical answer
to the question of whether one has here the Rota--Baxter framework.
For this is obviously the case when one deals only with logarithmic
divergences; and indeed most often only the latter is required. In
general, differentiation of an amplitude with respect to an external
momentum lowers the overall degree of divergence of a diagram. In~DR,
the Caswell--Kennedy theorem~\cite{CaswellK1982} states that the pole
part of any diagram, besides being independent of the scale, is a
polynomial in the external momentum. This follows easily from that
derivation and the projector $R_-$ commute in~DR. But even in the BPHZ
scheme $\partial_p t^l=t^{l-1}\partial_p$, and this is enough for the
differentiation trick to work.

Let us then consider the $J_F(p)\in \cS'_{s(F)}(\R^{4N})$
of~\eqref{eq:stone-of-contention}. Suppose moreover the multi-loop
divergent graph~$F$ has all its 1PI subgraphs~$\gamma$ made
convergent by application of Bogoliubov's preparation map~$\bar
R$. Then the renormalized integral $J^{\rm ren,BPHZ}_F(p)$ can be
defined as
\begin{equation}
 \label{eq:first-things-first}
    J^{\rm ren,BPHZ}_F(p) = \bigg[\int\prod_{l=1}^{|F|} d^4k_l\bigg]
     \big(I_F(p,k) - t^{s(F)}_p{\bar R}I_F(p,k)\big)
     =: \bigg[\int\prod_{l=1}^{|F|} d^4k_l\bigg]R_F(p,k).
\end{equation}
This recipe is however not unique. We can write as well
\begin{equation}
    J^{\rm ren,BPHZ}_F(p) = P^{s(F)}(p) + \bigg[\int\prod_{l=1}^{|F|}
    d^4k_l\bigg]R_F(p,k),
\label{eq:hammer-and-anvil}
\end{equation}
with $P^{s(F)}$ a polynomial of order~$s(F)$ in the external
momenta. This effects a `finite renormalization', in principle
undetermined, that might be put to use to fulfil renormalization
prescriptions (again, the form of that polynomial is severely
restricted by Lorentz invariance).

We now come to the key point.  The coefficients of~$P^{s(F)}$
in~\eqref{eq:hammer-and-anvil} exhibit the ambiguity of
renormalization in the context of the BPHZ scheme.  On the face of
it, the `pole terms' $t^{s(F)}_p I_F(p,k)$ do not depend at all on
the mentioned coefficients, and thus locality of the BWH
decomposition is guaranteed, in a trivial sense.  On the other
hand, the Galois group approach to renormalization
theory~\cite{cm2004,cm22004} stems originally from the idea that
ambiguities should be, insofar as possible, handled from a
group-theoretic point of view, much as classical Galois theory
handles the multiple solutions of polynomial equations.  Here
however the mentioned form of the ambiguity does not apparently
lend itself to RG~analysis.  We contend, however, that the
ambiguity is expressed essentially in the same form as before. The
Caswell--Kennedy theorem is suggestive of a direct link between
the DR and BPHZ formalisms, and next we endeavour to prove the
pertinence of the RG to BPHZ renormalization by the most direct
possible route: introducing a mass scale in the latter formalism
in direct analogy with the former.

To express the ambiguity implicit in the~$P^{s(F)}$
of~\eqref{eq:hammer-and-anvil} in terms of a mass scale, we use
the modified BPHZ scheme proposed in~\cite{ScheuBlume}.  For
instance, it is well known that the famous
graph~$\scalebox{0.85}{\fish}$ (`fish'~graph) giving the first
nontrivial contribution to the vertex correction in the~$\vf^4_4$
model in the Euclidean yields the amplitude
$$
     J^{\rm DR}_{\rm fish}(p) = \tilde{g}^2\mu^{2\eps} \int
     \frac{d^{\mathrm{D}}k}{(2\pi)^4}\,
     \frac{1}{k^2+m^2}\,\frac{1}{(p+k)^2+m^2},
$$
where $p=p_1+p_2$, say, and that, by use of Feynman's parametrization
(see below) and relation~\eqref{eq:tapa-del-perol} one obtains
\begin{equation*}
     J^{\rm DR}_{\rm fish}(p) = g\frac{\tilde{g}}{(4\pi)^2}
     \biggl[\frac{2}{\eps} + \int_0^1 dz\,
     \log\frac{{\overline\mu}^2}{p^2z(1-z) + m^2}\, + O(\eps)\biggr].
\end{equation*}
Now, the natural `zero point' for the mass scale in this problem is
clearly~$m$, and we note
$$
    R_+\big(J^{\rm DR}_{\rm fish}(p=0;\overline\mu=m)\big) = 0,
$$
as $\eps\downarrow0$. This, together with the mentioned
Caswell--Kennedy theorem, feeds the suspicion that the last expression
is just the $J^{\rm ren,BPHZ}_{\rm fish}(p)$
of~\eqref{eq:first-things-first}. The suspicion is correct. The
computation required for the renormalized fish~graph in the BPHZ
scheme is
\begin{equation}
    g^2\int\frac{d^4k}{(2\pi)^4}(1 - t^0_p) \bigg(\frac{1}{k^2 + m^2}
    \,\frac{1}{(p + k)^2 + m^2}\bigg).
\label{eq:hic-Rhodas}
\end{equation}
Introduce the Feynman trick, prior to the Taylor subtraction,
\begin{align}
    &g^2\int_0^1 dz\int\frac{d^4k}{(2\pi)^4}(1 - t^0_p)\,
    \frac{1}{\big[((p + k)^2 + m^2)z + (1 - z)(k^2 + m^2)\big]^2}
    \nonumber \\
    &= g^2\int_0^1 dz\int\frac{d^4k}{(2\pi)^4}(1 - t^0_p)
    \frac{1}{[k^2 + p^2z(1 - z) + m^2]^2}.
\label{eq:fishandchips}
\end{align}
The translation $k\to k-zp$, depending on the Feynman parameter, has
been made in order to obtain here the same denominator as in DR
calculations. With~$\Omega_4$ the area of the unit sphere in~$\R^4$,
the integral~\eqref{eq:fishandchips} now becomes
\begin{align*}
    &\frac{\Omega_4\,g^2}{(2\pi)^4}\int_0^1 dz\int_0^\infty dk\,
    \bigg[\frac{k^3}{[k^2 + p^2z(1 - z) + m^2]^2} - \frac{k^3}{[k^2 +
    m^2]^2}\bigg] \\
    &= \frac{g^2}{16\pi^2}\int_0^1 dz\,\log\frac{m^2}{p^2z(1 - z) + m^2}.
\end{align*}
The last step is to convert the $p$-independent part in the argument
of the logarithm into a mass scale: $m\to\overline\mu$.  With this, we
recover on the nose the DR result, in the ${\rm\overline{MS}}$~scheme
as it turns out.  Incidentally, as remarked in~\cite{Rio1995}, this
procedure allows us to give the exact value of the BPHZ
integral~\eqref{eq:hic-Rhodas}: the expression $\int_0^1
dz\,\log\big(1 + \frac{p^2}{m^2}z(1 - z)\big)$ is actually well known
in statistical physics, and leads by elementary manipulations
involving the golden ratio to
\begin{equation*}
    J^{\rm ren,BPHZ}_{\rm fish}(p) = -\frac{g^2}{16\pi^2}
    \bigg(\sqrt{1 + \frac{4m^2}{p^2}}\log\frac{\sqrt{4m^2/p^2} + 1}
    {\sqrt{4m^2/p^2} - 1} - 2\bigg).
\end{equation*}
Thus, what we have done above amounts to \textit{identify the
constant} term ---recall $P^{s(F)}(p)$
in~\eqref{eq:hammer-and-anvil}. We have the right to add to the
previous expression the term $g^2/16\pi^2$ times
$\log\big({\overline\mu}^2/m^2\big)$. We note also that one can
recover the residue~$g^2/8\pi^2$ here from~$J^{\rm ren,BPHZ}_{\rm
fish}$, as the coefficient of the term logarithmic in the scaling
factor,
$$
    J^{\rm ren,BPHZ}_{\rm fish}(\lambda p) \sim J^{\rm ren,BPHZ}_{\rm
    fish}(p) - \frac{g^2}{8\pi^2}\log\lambda,
$$
as~$\lambda\uparrow\infty$.

The steps of the modified BPHZ procedure are: (i)~Introduction of the
Feynman parame\-trization in~$J^{\rm BPHZ}_F(k,p)$.  (ii)~Exchange of
the integrations.  (iii)~Translation of the integration variables
by~$\lambda p$, with~$\lambda$ dependent on the Feynman parameter.
(iv)~Taylor subtraction.  (v)~Integration over loops and replacement
of the mass~$m$ in the $p$-constant part of the resulting logarithm by
a mass scale.  There is nothing to forbid the same operations to be
performed on any primitive logarithmically divergent graph of any
field theory and then we are optimistic that, by use of skeletal
expansions and the integral equations, we would be led to a procedure
largely parallel to~DR, and so to a brute-force proof that the
coefficients of the higher powers of the scaling logarithms in BPHZ
renormalization are determined by the residues.  To verify this with
full particulars, however, would take us too far afield.

Recapitulating, the Lie-theoretic method shows promise in dealing
with renormalization schemes others than~DR with~MS. The presence
of a Rota--Baxter structure is a requisite for the validity of
such a framework; it obviously holds for the
${\rm\overline{MS}}$~prescription in~DR as well.  What of other
renormalization methods?  For massive fields, the BPHZ scheme does
verify the required conditions.  We have learned, however, that
the details are very idiosyncratic: as stated above, locality is
moot; and the beta function enters the picture through
Theorem~\ref{thm:prueba-de-fuego}, referring to renormalized
quantities, rather than to counterterms.  For massless fields, the
price of the Rota--Baxter property is relinquishing Lorentz
invariance, and this is too heavy to pay. The Taylor subtraction
in Epstein--Glaser renormalization has roughly the same properties
as the BPHZ scheme, both in regard of massive and massless fields;
nevertheless, there one is confronted to the problems of good
definition that plague the attempts~\cite{BK,Etoile}. Procedures
based on analytic renormalization or Hadamard regularization
\cite{jmgb2003} have not been investigated yet from the
Rota--Baxter viewpoint.  Thus it is too early in the game to draw
a list of known schemes that would definitely fit in our approach;
we plan to come to this in future work.  It is intriguing that the
case study of BPHZ renormalization points out to the pertinence of
the ${\rm\overline{MS}}$~prescription in~DR.


\section{On Connes--Marcolli's motivic Galois theory}
\label{sect:cosmic}

In the Connes--Kreimer picture of renormalization, group theory
and the scheme-theoretic equivalent theory of commutative Hopf
algebras have become a fundamental tool of pQFT. Connes and
Marcolli identified recently~\cite{cm2004} a new level at which
Hopf algebra structures enter pQFT. Namely, they constructed an
affine group scheme $U^\ast$, universal with respect to physical
theories, and pointed out its analogies with number theory, e.g.
with the motivic Galois group of the scheme of 4-cyclotomic
integers $\Z[i][{\frac{1}{2}}]$.

In their work the initial physical problem attacked through the
Connes--Kreimer paradigm translates into the classification of
equisingular $G$-valued flat connections on the total space of a
principal bundle over an infinitesimal punctured disk (with $G$ the
group scheme represented by~$H$). From the representation theoretic
point of view, the classification is provided by representations
$U^\ast\longrightarrow G^\ast$, where $U^\ast$ is the semi-direct
product with the grading of the pro-unipotent group~$U$, the Lie
algebra of which is the free graded Lie algebra with one generator
$e_n$ in each degree $n>0$, and similarly for~$G^\ast$. Returning to the
geometrical side of the correspondence and featuring the DR setting
that leads to the Riemann--Hilbert picture of renormalization, Connes
and Marcolli construct a universal singular frame on principal
$U$-bundles over~$B$. A formal expression for it is given by:
$$
    \gamma (\varepsilon,v)=\sum\limits_{n\geq 0}\sum\limits_{k_j}
    \frac{e(k_1)\cdots e(k_n)}{k_1(k_1 + k_2)\cdots (k_1 + \cdots +
    k_n)}v^{\sum k_j}\varepsilon^{-n}.
$$
As already remarked in~\cite{cm2004} and our introduction, it is
interesting that the coefficients of the frame are essentially those
appearing in the index formula of Connes--Moscovici; this would hint
at the rooting of noncommutative geometry in quantum field theory,
which has been Connes' contention for a long while.

We have already shown that other Hopf algebra structures (or, from the
scheme-theoretic point of view, pro-unipotent groups) do appear
naturally in pQFT, namely the Hopf algebras $\Char(A)$ of
characteristic functions associated to commutative target algebras,
e.g., although not exclusively, of quantum amplitudes.  These Hopf
algebra structures arise naturally from algebraic-combinatorial
constructions on Hopf algebras, and therefore do not immediately
relate to the geometrical-arithmetical constructions underlying the
definition of the motivic Galois group in~\cite{cm2004}.
Nevertheless, the formula rendering the universal singular frame in
the motivic understanding of renormalization also essentially
coincides with our map~$\Gamma$ ---the inverse of the Dynkin map.
This indicates that the practical consequences for renormalization of
the Riemann-Hilbert and/or motivic point of view can be translated to
the setting of FLA theory ---which, besides being more familiar to
many, is independent of the geometry embedded in the DR~scheme.  As it
turns out, the pro-unipotent groups/Hopf algebras $\Char(H)$ and
$\Char(A)$ are related naturally to the group~$U$.  In the remainder
of the present section, we would like to make explicit how both
viewpoints connect ---although the reasons behind this connection
certainly ought to be deepened.

Let us recall a few general facts from the theory of Solomon algebras
---see~\cite{patras1994,reutenauer1993} for details.  Let~$\sigma$ be
a permutation in the symmetric group~$S_n$ of order~$n$.  The
descent set $D(\sigma )$ of~$\sigma$ is the set $D(\sigma):=
\{i,\,\sigma (i) > \sigma (i+1)\}$. Note $n \notin D(\sigma)$. The
descent composition $C(\sigma)$ of~$\sigma$ is the composition
of~$n$ (that is, the sequence $(c_1,\ldots,c_k)$ of strictly
positive integers of total sum~$n$) such that, when viewed as a
word, $\sigma=\sigma(1)\ldots\sigma (n)$ can be written $u_1\ldots
u_k$, where each word~$u_i$ is increasing and of length~$c_i$, and
where~$k$ is minimal.  For example, $D(21534)=\{1,3\}$ and
$C(21534)=(1,2,2)$. The notions of descent set and descent
composition are obviously equivalent, the equivalence being
induced by the map:
$$
(c_1,\ldots,c_k) \longmapsto \{c_1,c_1 + c_2,\ldots,c_1 + \cdots +
c_{k-1}\}.
$$
The Solomon algebra~$\Sigma_n$ of type $A_n$ was first introduced
as a \textit{noncommutative lift} to the group algebra of the
representation ring of $S_n$~\cite{solomon1976}.  As a vector
space, $\Sigma_n$ is the linear span of the elements $D_{\subseteq
S}$ in~$\Q[S_n]$, where~$S$ runs over subsets of~$[n-1]$ and
$$
D_{\subseteq S} := \sum_{\substack{\sigma\in S_n \\
D(\sigma)\subseteq S}}\sigma.
$$
Then Solomon's fundamental theorem states that $\Sigma_n$ is closed
under the composition product in~$S_n$.

Now, let $X$ be an infinite alphabet.  The dual graded Hopf
algebra $T^\ast(X)=\bigoplus_{n\in\N}T_n^\ast (X)$ of~$T(X)$ is
graded connected commutative, with the shuffle product as the
algebra product and the deconcatenation coproduct:
$$
x_{i_1}\ldots x_{i_n} \longmapsto \sum\limits_{k=0}^n x_{i_1}\ldots
x_{i_k}\otimes x_{i_{k+1}}\ldots x_{i_n},.
$$
where we view $x_{i_1} \ldots x_{i_n}$ as an element of $T^\ast
(X)$ using the usual pairing $\langle x_{i_1} \ldots
x_{i_n}|x_{j_1}\ldots x_{j_k} \rangle = \delta_{x_{i_1}\ldots
x_{i_n}}^{x_{j_1}\ldots x_{j_k}}$. The symmetric group of order
$n$ embeds into $\End(T_n^{\ast }(X))\subset\End(T^\ast(X))$:
$$
\sigma (x_{i_1}\ldots x_{i_n}) := x_{i_{\sigma^{-1}(1)}}\ldots
x_{i_{\sigma^{-1}(n)}}.
$$
This map induces an embedding of algebras of $\Sigma_n$ into
$\End(T^\ast(X))$, where the product on the latter algebra is the
composition of maps.

Let us write now $\mathcal D$ for the descent algebra of $T^\ast
(X)$, that is the convolution subalgebra of $\End(T^\ast (X))$
generated by the projections $p_n:T^\ast (X)\longrightarrow
T_n^\ast(X)$ on the graded components of $T^\ast (X)$. The algebra
$\mathcal D$ is naturally graded and provided with a Hopf algebra
structure for which the $p_n$ form a sequence of divided powers:
$$
\Delta (p_n) = \sum\limits_{i+j=n}p_i\otimes p_j.
$$
We write ${\mathcal D}_n$ for the component of degree $n$.

\begin{lem}
The convolution algebra $\mathcal D$ is also closed under the
composition of maps $\circ$ in $End(T^\ast(X))$.
\end{lem}

The result follows from Corollary~9.4 in \cite{reutenauer1993}
(where the dual setting is considered, that is, the convolution
subalgebra $\Gamma$ of $End(T(X))$ generated by the graded
projections in $T(X)$) and also follows directly from~\cite[Thm
II.7]{patras1994}.

\begin{prop}
The embedding of $\Sigma_n$ into $(End(T^\ast(X)),\circ )$ induces an
isomorphism of algebras
$$
\Sigma_n \longrightarrow {\mathcal D}_n.
$$
\end{prop}

The proof follows from Corollary~9.2 in \cite{reutenauer1993} by
duality.  It basically amounts to observe that, if $C(\sigma)=
(c_1,\ldots,c_k)$, then $\sigma^{-1}$ is a
$(c_1,\ldots,c_k)$-shuffle. For example, if $\sigma=(21534)$ then
$C(\sigma )=(1,2,2)$ and $\sigma^{-1}=(21453)$, so that the word
$\sigma (x_1...x_5)=x_2x_1x_4x_5x_3$ is a shuffle of $x_1$,
$x_2x_3$ and $x_4x_5$, and appears therefore in the expansion of
$p_1\ast p_2\ast p_2 (x_1...x_5)$.

\begin{prop}
The algebra $\mathcal D$ is freely generated as an associative
algebra by the graded projections $p_n$. Equivalently, it is
freely generated by the graded components $p_n\circ D$ of the
Dynkin idempotent $D=S\ast Y$, regarded as an element of
$\End(T^\ast(X))$.
\end{prop}

The first part of the proof of this proposition is Corollary~9.14
of~\cite{reutenauer1993} (stated for $\Gamma$, that is, in the
dual setting). The second assertion is found e.g.
in~\cite[Sect.~5.2]{gelfand1995}.

\begin{cor}
Regarded as a pro-unipotent group scheme, the graded dual Hopf
algebra $\mathcal D^\ast$ is canonically isomorphic to the ring of
coordinates of the Connes--Marcolli group~$U$ of renormalization
theory.
\end{cor}

Through this correspondence, by our Lemma~\ref{dynkid}, the
coefficients of the universal singular frame are reflected in the
coefficients of the expansion of the identity of $T^\ast (X)$ on
the natural linear basis of $\mathcal D$ viewed as the free
associative algebra generated by the graded components of the
Dynkin operator.  Now, the universal properties of the Galois
group $U$ for renormalization, when the group is understood by
means of $\mathcal D$, follow from the constructions
in~\cite{patras1994}, where it is shown that the descent algebra
is an algebra of natural (endo)transformations of the forgetful
functor from graded connected commutative Hopf algebras to graded
vector spaces ---that is, an universal endomorphism algebra for
graded connected commutative Hopf algebras. In other terms, there
is a natural map from $\mathcal D$ to $\End(H)$, where $H$ is an
arbitrary graded connected commutative Hopf algebra.  Using the
arguments developed in the first sections of this article, one
shows easily that this map factorizes through an Hopf algebra map
from $\mathcal D$ to~$\Char(H)$; this follows e.g. from the fact
that the graded projections generate $\mathcal D$ as a convolution
algebra and form a sequence of divided powers both in ${\mathcal
D}\subset\End(T^\ast (X))$ and in~$\Char(H)$.  In summary,

\begin{cor}
The descent algebra $\mathcal D$ acts naturally by right composition
on $\Lin(H,A)$.  Moreover, the group of group-like elements in
$\mathcal D$ acts naturally on the group $G(A)$ of Feynman rules.
\end{cor}

The second part of the corollary follows from the third identity in
Lemma~\ref{lem:AstCirc}.

Besides providing a natural link between the Galoisian approach to
renormalization and the noncommutative representation theory of
the symmetric groups~\cite{BlessenohlS}, the combinatorial
approach implies moreover that the Connes--Marcolli universal
Galois group~$U$ inherits the very rich structure of the descent
algebra. The appearance of the descent algebra (or equivalently of
the Hopf algebras of noncommutative symmetric functions and
quasi-symmetric functions, see \cite{gelfand1995}), beyond
broadening the scope of the mathematical theory of
renormalization, should result into new developments in the field,
possibly complementary with the arithmetic ones.


\vspace{0.4cm}

\textbf{Acknowledgements}

\smallskip

The first named author\footnote{currently at: Max Planck Institute for
Mathematics, Vivatsgasse 7, D-53111 Bonn, Germany} acknowledges
greatly the support by the European Post-Doctoral Institute and the
Institut des Hautes \'Etudes Scientifiques (I.H.\'E.S.).  He is also
indebted to Laboratoire J.~A.~Dieudonn\'e at Universit\'e de Nice
Sophia-Antipolis for warm hospitality.  He is very grateful to
C.~Bergbauer for useful discussions.  The second named author
acknowledges partial support from CICyT, Spain, through
grant~FIS2005-02309.  He is also grateful for the hospitality of
Laboratoire J.~A.~Dieudonn\'e.  The present work received support from
the ANR grant AHBE~05-42234.  We are pleased to thank the anonymous
referee, whose comments prompted us to clarify several aspects of the
paper.


\end{document}